\documentclass[a4paper,epj,nopacs]{svjour}
\usepackage{amsfonts,amssymb,latexsym}
\usepackage{graphicx}
\usepackage[british]{babel}
\usepackage{epsfig}
\newlength{\colw}
\begin{document}
\title{Numerical simulation tests with light dynamical quarks}
\author{
The qq+q Collaboration \\[4mm]
Federico Farchioni\inst{1}, Claus Gebert\inst{2}, Istv\'an Montvay\inst{2} \and 
Luigi Scorzato\inst{2}
}                     
%
%
\institute{
         Institut f\"ur Theoretische Physik, Universit\"at M\"unster,
         Wilhelm-Klemm-Str. 9, D-48149 M\"unster, Germany
         \and 
         Deutsches Elektronen-Synchrotron DESY,
         Notkestr.\,85, D-22603 Hamburg, Germany
}
\date{June 2002, revised August 2002}
%
\abstract{
Two degenerate flavours of quarks are simulated with small masses
 down to about one fifth of the strange quark mass by using the
 two-step multi-boson (TSMB) algorithm.
 The lattice size is $8^3 \cdot 16$ with lattice spacing about
 $a \simeq 0.27\,{\rm fm}$ which is not far from the $N_t=4$
 thermodynamical cross-over line.
 Autocorrelations of different physical quantities are estimated
 as a function of the quark mass.
 The eigenvalue spectra of the Wilson-Dirac operator are investigated.
\PACS{
      {PACS-key}{discribing text of that key}   \and
      {PACS-key}{discribing text of that key}
     } 
} 
\authorrunning{Federico Farchioni et al.}
\titlerunning{Numerical simulation tests with light dynamical quarks}
\newcommand{\be}{\begin{equation}}
\newcommand{\ee}{\end{equation}}
\newcommand{\half}{\frac{1}{2}}
\newcommand{\rar}{\rightarrow}
\newcommand{\lar}{\leftarrow}
\newcommand{\LCB}{\raisebox{-0.3ex}{\mbox{\LARGE$\left\{\right.$}}}
\newcommand{\RCB}{\raisebox{-0.3ex}{\mbox{\LARGE$\left.\right\}$}}}
\newcommand{\U}{\mathrm{U}}
\newcommand{\SU}{\mathrm{SU}}
%
\maketitle
\section{Introduction}\label{sec1}

 The question about the computational cost of dynamical quark
 simulations is a central issue in lattice gauge theory.
 Existing unquenched simulations are typically done in a region where
 the quarks are not light enough, in most cases -- especially in case
 of Wilson-type quarks -- with two light quark flavours ($u$ and $d$)
 having masses larger than half of the strange quark mass
 ($m_{ud} > \half m_s$).
 The physical masses of $u$- and $d$-quarks are so small that in the
 foreseeable future simulations can only be carried out at somewhat
 higher masses.
 In order to extrapolate the results to the physical masses chiral
 perturbation theory based on the low energy chiral effective Lagrangian
 can be used.
 However the systematic errors can only be controlled if the
 dynamical quark masses in the simulations are close enough to the
 physical point.
 For instance, in case of partially quenched simulations to determine
 the low energy constants in the chiral effective Lagrangian of QCD
 we would like to reach at least $m_{ud} \leq \frac{1}{4} m_s$
 \cite{Sharpe:SHORESH}.

 Going to light quark masses in unquenched QCD simulations is a great
 challenge for computations because known algorithms have a
 substantial slowing down towards small quark masses.
 The present status has been recently summarized by the contributors to
 the panel discussion at the Berlin lattice conference
 \cite{Christ:BERLIN,Gottlieb:BERLIN,Jansen:BERLIN,Lippert:BERLIN,Ukawa:BERLIN,Wittig:BERLIN}.
 Inspired by the results presented there the computational cost of a
 simulation with two light quarks will be parametrized in the present
 paper as
\begin{equation}\label{eq1.01}
C = F\; (r_0 m_\pi)^{-z_\pi} \left(\frac{L}{a}\right)^{z_L}
\left(\frac{r_0}{a}\right)^{z_a} \ .
\end{equation}
 Here $r_0$ is a physical length, for instance the Sommer scale parameter
 \cite{Sommer:SCALE}, $m_\pi$ the pion mass, $L$ the lattice extension and
 $a$ the lattice spacing.
 The powers $z_{\pi,L,a}$ and the overall constant $F$ are empirically
 determined.
 The value of the constant factor $F$ depends on the precise definition
 of ``cost'' \cite{Frezzotti:BENCHMARK}.
 For instance, one can consider the number of floating point operations
 in one autocorrelation length of some important quantity, or the
 number of fermion-matrix-vector-multiplications necessary for 
 achieving a given error of a quantity.
 Of course, the cost also depends on the particular choice of lattice
 action and of the dynamical fermion algorithm which should be
 optimized.

 An alternative parametrization can be obtained from the one in
 (\ref{eq1.01}) by replacing the powers of $r_0 m_\pi$ by those of
 $m_\pi/m_\rho$.
 In fact, the results of the CP-PACS, JLQCD Collaboration have been
 presented by A.~Ukawa at the Berlin lattice conference
 \cite{Ukawa:BERLIN} in this form:
\begin{eqnarray} \label{eq1.02}
& & C_U = F_U\; \left(\frac{m_\pi}{m_\rho}\right)^{-z_{\pi\rho}}
\left(\frac{L}{a}\right)^{z_L} \left(\frac{r_0}{a}\right)^{z_a} \ ,
 \\[0.7em]
& & F_U = 5.9 \cdot 10^6\, {\rm flop} \\
& & z_{\pi\rho} = 6 \ , \hspace*{1em}
z_L = 5 \ , \hspace*{1em}
z_a = 2 \ .
\end{eqnarray}
 Since the determination of the $\rho$ meson mass is difficult for
 light quarks when the decay $\rho \to \pi\pi$ is allowed, we prefer
 the form in (\ref{eq1.01}).
 Other parametrizations used for Wilson-type quarks
 \cite{Lippert:BERLIN,Wittig:BERLIN} are given under the assumption
 that $z_\pi=z_a \equiv z_{a\pi}$ when in (\ref{eq1.01}) the physical
 length parameter $r_0$ disappears.

 In the present paper we report on the results of extended test runs
 with the simple Wilson fermion action using the two-step multi-boson
 algorithm \cite{Montvay:TSMB} in order to determine the quark mass
 dependence of the computational cost of dynamical Monte Carlo
 simulations with two light flavours in the region
 $m_{ud} \geq \frac{1}{5} m_s$.
 For the definition of the quark mass the dimensionless quantity
\begin{equation}\label{eq1.03}
M_r \equiv (r_0 m_\pi)^2
\end{equation}
 is used, which already appears in (\ref{eq1.01}).
 This is a possible definition for small quark masses because
 for $m_q \to 0$ the pion mass behaves as $m_\pi \propto \sqrt{m_q}$.
 For defining the value of $M_r$ which corresponds to the strange quark
 mass one can use unquenched $N_f=2$ lattice data.
 For instance, the experimental value of the $\Omega^-$ baryon mass
 $m_{\Omega^-} = 1.672\,{\rm GeV}$ and $r_0 = 0.5\, {\rm fm}$
 give $r_0 m_{\Omega^-} = 4.237$.
 Interpolating CP-PACS results \cite{AliKhan:NF2SPECT} for the $\Delta$
 baryon mass at their largest $\beta$ value $\beta=2.20$ between
 $\kappa=0.1363$ and $\kappa=0.1368$ one can match 
 $r_0 m_\Delta = 4.237$ if their pion mass is
 $r_0 m_\pi \simeq 1.76$.
 This gives for the strange quark mass $M_{r,strange} \simeq 3.1$.
 Of course, there are also other ways to estimate $M_{r,strange}$
 which might give slightly different values.
 In the present paper, without attempting to really compute the
 strange quark mass, we shall stick to the operational definition
\begin{equation}\label{eq1.04}
M_{r,strange} \equiv 3.1 \ .
\end{equation}

 The Monte Carlo simulations are done near the $N_t=4$ thermodynamical
 cross-over line that is for $a \simeq 0.27\,{\rm fm}$.
 The lattice size is $8^3 \cdot 16$ implying a physical lattice
 extension $L \simeq 2.2\,{\rm fm}$.
 Later on we shall also extend our investigations to $12^3 \cdot 24$
 and $16^3 \cdot 32$ lattices.
 Our present studies can be considered as complementary to the ones on
 larger lattices (closer to the continuum limit) but at larger quark
 masses (typically $m_{ud} \geq \half m_s$) 
 \cite{Christ:BERLIN}-\cite{Wittig:BERLIN}.

 In addition to obtaining estimates of autocorrelation lengths as a
 function of the quark mass we also performed a detailed study of the
 small eigenvalue spectra both for the hermitean and non-hermitean
 Wilson-Dirac fermion matrix.
 Besides giving important qualitative information about quark dynamics
 this also allows to clear the issue of the {\em sign problem} of the
 quark determinant.
 For an odd number of Wilson-type quark flavours the fermion determinant
 can have both signs because there might be some eigenvalues (of the
 non-hermitean fermion matrix) on the negative real axis.
 Since for importance sampling a positive measure is required, the
 determinant sign can only be taken into account in a measurement
 reweighting step.
 A strongly fluctuating determinant sign is a potential danger for the
 effectiveness of the Monte Carlo simulation because cancellations
 can occur resulting in an unacceptable increase of statistical errors.
 We actually study this question here with two degenerate quark flavours
 ($N_f=2$) where in the path integral the square of the fermion
 determinant appears and hence the sign is irrelevant.
 But our two quarks are much lighter than the physical $s$-quark.
 Therefore the statistical insignificance of negative eigenvalues in
 this case hints towards the absence of the sign problem in the physical
 case of  $N_f=2+1$ quark flavours, when the sign of the $s$-quark
 determinant could, in principle, cause a problem.

 The plan of this paper is as follows: in the next section we shortly
 introduce the parameters of the TSMB algorithm and give some details
 of our implementation on different computers.
 In section \ref{sec3} the autocorrelations are investigated for some
 basic quantities such as the average plaquette and the pion mass.
 Section \ref{sec4} contains a detailed study of the small eigenvalue
 spectra of the fermion matrix.
 The last section is devoted to discussion and conclusions.

\section{The TSMB algorithm}\label{sec2}

 We use in this study the two-step multi-boson (TSMB) algorithm which
 has been originally developed for Monte Carlo simulations of the
 supersymmetric Yang-Mills theory \cite{Montvay:TSMB} but it can also
 be applied more generally \cite{Montvay:QCD}.

\subsection{Algorithmic parameters}\label{sec2.1}

 TSMB is based on a representation of the fermion determinant in the
 form
\be\label{eq2.01}
\left|\det(Q)\right|^{N_f} \;\simeq\;
\frac{1}{\det P^{(1)}_{n_1}(\tilde{Q}^2)
\det P^{(2)}_{n_2}(\tilde{Q}^2)} \ .
\ee
 Here $N_f$ denotes the number of fermion flavours and $Q$ is the
 fermion matrix, which in the present paper is equal to the
 Wilson-Dirac matrix
\begin{eqnarray}  \label{eq2.02}
Q_{ys,xr}   \equiv 
\delta_{yx}\delta_{sr} - \ \ \ \ \ \ \ \ \ \ \ \ \ \ \ \ \ \ \ \ \\ \nonumber
\kappa \sum_{\mu=1}^4 \left[ 
\delta_{y,x+\hat{\mu}}(1+\gamma_\mu) U_{sr,x\mu} +
\delta_{y+\hat{\mu},x}(1-\gamma_\mu) U^\dagger_{sr,y\mu}  \right]
\end{eqnarray}
 with $x,y$ denoting lattice sites, $r,s$ colour (triplet) indices,
 $\hat{\mu}$ the unit lattice vector in direction $\mu$,
 $U_{x\mu} \in {\rm SU}(3)$ gauge link matrices and $\kappa$ the
 hopping parameter.
 The hermitean Wilson-Dirac fermion matrix is defined as usual by
\be \label{eq2.03}
\tilde{Q} \equiv \gamma_5 Q = \tilde{Q}^\dagger \ . 
\ee
 The polynomial approximations in (\ref{eq2.01}) satisfy
\begin{eqnarray} \label{eq2.04}
P^{(1)}_{n_1}(x) &\simeq& x^{-N_f/2} \ ,
\nonumber \\[0.7em]
\lim_{n_2 \to \infty} P^{(1)}_{n_1}(x)P^{(2)}_{n_2}(x) &=&
x^{-N_f/2} \ , \hspace{2em}
x \in [\epsilon,\lambda]
\end{eqnarray}
 where the interval $[\epsilon,\lambda]$ covers the spectrum of the
 squared hermitean fermion matrix $\tilde{Q}^2$ on a typical gauge
 configuration.
 The first polynomial $P^{(1)}$ is a crude approximation with relatively
 low order.
 It is used in the multi-boson representation of fermion determinants
 \cite{Luscher:MB}.
 The second polynomial $P^{(2)}$ is a correction factor which is taken
 into account in the gauge field updating by a global accept-reject
 step.
 For this a polynomial approximation of the inverse square root of
 $P^{(2)}$ is also needed:
\be \label{eq2.05}
 P^{(3)}_{n_3}(x) \simeq P^{(2)}_{n_2}(x)^{-\half} \ .
\ee
 The limit $n_2 \to \infty$ can be taken in the computed expectation
 values if one produces several update sequences with increasing $n_2$
 or, more conveniently, one can keep $n_2$ fixed at some
 sufficiently large value for a good approximation and introduce a
 further polynomial $P^{(4)}$ satisfying
\be\label{eq2.06}
\lim_{n_4 \to \infty} P^{(1)}_{n_1}(x)P^{(2)}_{n_2}(x)P^{(4)}_{n_4}(x)
= x^{-N_f/2} \ .
\ee
 $P^{(4)}$ can be taken into account by reweighting the gauge
 configurations during the evaluation of expectation values.
 In most cases the order $n_2$ of $P^{(2)}_{n_2}$ can be chosen high
 enough such that the reweighting correction has a negligible effect on
 expectation values.
 In any case the evaluation of the reweighting factors is useful because
 it shows whether or not the two-step approximation in (\ref{eq2.04}) is
 good enough.
 For a recent summary of some details of TSMB and for references see
 section 3 of \cite{Montvay:SYMREV}.

 The Monte Carlo integration of the path integral is performed by
 averaging over a sequence (Markov chain) of multi-boson and gauge field
 configurations.
 The $n_1$ multi-boson fields ($\Phi$) and gauge fields ($U$) are
 updated in repeated {\em update cycles} consisting of several
 sweeps over the multi-boson fields and gauge field.
 For the multi-boson fields we use (local) heatbath and overrelaxation
 as well as global quasi-heatbath \cite{deForcrand:QUASIHEAT} sweeps.
 For the gauge field update heatbath and overrelaxation sweeps are
 alternated.
 After several gauge field sweeps a global Metropolis accept-reject 
 correction step is performed by the polynomials $P^{(2)}$ and $P^{(3)}$.
 The update sequence within a cycle is subject to optimization with the
 goal to decrease autocorrelations.
 We tried several kinds of update sequences within an update cycle.
 A typical sequence was: 3 $\Phi$-overrelaxations, 1 $\Phi$-heatbath,
 12 $U$-overrelaxation, global $U$-Metropolis, 3 $\Phi$-overrelaxations,
 1 $\Phi$-heatbath, 6 $U$-heatbath, global $U$-Metropolis.
 In every 10-th cycle the first $\Phi$-overrelaxation-$\Phi$-heatbath
 combination was replaced by a global quasi-heatbath.

\subsection{Implementation and performance}\label{sec2.2}

 We have implementations of the updating and measurement programs in
 TAOmille for the APEmille and in C++/ MPI. The latter implementation
 is usable on many different architectures as long as they provide 
 a C++ compiler and, in case of parallel computers, support MPI.
 In the updating program the computing time is dominated by the
 fermion-matrix-vector-multiplications (MVMs), $2\cdot(n_2+n_3)$ of them
 are needed for the correction step and ${\cal O}(100 \cdot n_1)$ for the
 global heatbath and quasi-heatbath \cite{deForcrand:QUASIHEAT}.
 Altogether they make up $60\%-80\%$ of the computing time.
 In the most interesting regions of small quark masses the program is
 dominated by the MVMs even more strongly.
 The same is true for the measurement program, where smearing and
 calculation of simple Wilson loops takes only a few percent of the time.
 It is therefore of the utmost importance to improve the performance of the MVM
 routines, both preconditioned (for the correction step and the
 measurements) and non-preconditioned (for the global heatbath).
 This has been done for the APEmille, the Cray T3E with the KAI C++
 compiler, and for a multi-node Pentium-4 cluster here also exploiting the
 possibilities of SSE and SSE2 instructions. 
 Results are given in table \ref{tab_perf}.
 Note that an important feature of the SSE instructions is that in
 single precision the peak performance is doubled compared to double
 precision.
 The performance numbers in table \ref{tab_perf} are substantially
 influenced by the communication costs among computing nodes.
 Without communications the numbers both for APEmille and P4-cluster
 would be almost a factor of two higher.
 On larger volumes than those considered here communication will have less
 influence on the performance.

\begin{table}
\caption{
 Performance of the matrix-vector-multiplication in MFlops and percent 
 relative to peak performance on one board (8 nodes) on the APEmille 
 and on 8 processors on the T3E and P4-cluster for a $8^3 \cdot 16$
 lattice.}
\label{tab_perf}
\begin{tabular}{|c|c|c|c|}
\hline
        & APEmille      &\hspace{1em} T3E-1200\hspace{1em} & P4-1700 
\\ \hline
 32 bit & 1008 (23.9\%) & 912 (9.5\%) & 4322 (7.9\%)   \\ \hline
 64 bit &               & 712 (7.4\%) & 2087 (7.7\%)   \\ \hline
\end{tabular}
\end{table}

 Since the matrix multiplications dominate the computing time it
 is reasonable to express e.g. autocorrelations in units of MVMs.
 The remaining part of the computation is given by the local updates.
 These are composed of parts which can be  essentially thought of
 as pieces of MVMs, too. As a result the following approximate formula 
 for the total amount of MVMs needed for one update cycle is obtained:
\begin{equation} \label{eq2.07}
N_{MVM}/\mbox{cycle} \;\simeq\; 
6\,(n_1N_\Phi+N_U)+2\,(n_2+n_3)N_{C}+I_G F_G \ .
\end{equation}
 Here $N_\Phi$ is the number of local bosonic sweeps per update cycle, 
 $N_U$ the number of local gauge sweeps, $N_{C}$ the number 
 global Metropolis accept-reject correction steps, and $I_G$ and $F_G$ give 
 the number of MVMs and frequency of the global heatbath.

 For data from APEmille and Cray 
 the estimate of the cost of the local updates obtained from (\ref{eq2.07})
 agrees with the actual costs up to $5\%$. Therefore the 
 final costs in units of MVM based on (\ref{eq2.07})
 are not much influenced by the approximation.
 This is not true for the data presented for the P4-1700 system,
 since in this case the matrix multiplication and the local updates
 are not treated homogeneously. Indeed the former 
 is written in assembler using  SSE/SSE2 instructions while 
 our code for the local updates is written in C++ and compiled with 
 the g++ compiler.
 As a result, the estimate for the cost of the local updates is in this
 case underestimated by about a factor three.
 Still we take the above formula as a reference when tuning the
 parameters because the number of MVMs is more generally applicable
 as it does not depend on implementation details.
 In addition, in the future the local updates could be rewritten
 by using SSE/SSE2 instructions, too, so eliminating the non-homogeneity 
 with the MVMs.

 It is sometimes interesting to convert the number of MVMs into
 the number of floating point operations.
 On our $8^3 \cdot 16$ lattice this conversion is approximately
\begin{equation} \label{eq2.08}
1\; {\rm MVM} \;\simeq\; 1.1 \cdot 10^7\; {\rm flop} \ .
\end{equation}
%

\section{Autocorrelations at small quark masses}\label{sec3}

 The bare parameters of the QCD lattice action with Wilson quarks
 ($\beta$ for the SU(3) gauge coupling and $\kappa$ for the
 hopping parameter of two degenerate quarks) have to be tuned properly
 in order to obtain the desired parameters in the Monte Carlo
 simulations.
 We are interested in the quark mass dependence of the simulation cost
 of hadron spectroscopy applications, therefore we want to keep the
 physical volume of our lattices sufficiently large and (approximately)
 constant.
 For a $8^3 \cdot 16$ lattice, a lattice spacing
 $a \simeq 0.27\,{\rm fm}$ implies a lattice extension of
 $L \simeq 2.2\,{\rm fm}$ which is a reasonable starting point for
 spectroscopy.
 Previous Monte Carlo simulations with $N_f=2$ Wilson quarks
 \cite{Iwasaki:NT4,Blum:NT6} showed that this kind of lattice
 spacing is realized near the $N_t=4$ and $N_t=6$ thermodynamical
 transition lines which, therefore, provide a good orientation.
 We started our simulations at a relatively large quark mass on the
 $N_t=4$ transition line and then tuned $\beta$ and $\kappa$ towards
 smaller quark masses keeping $r_0/a$ approximately constant.
 A summary of simulation points is given in table \ref{tab_run}
 where some important algorithmic parameters of the TSMB are also collected.
\begin{table*}
\caption{
 Bare couplings, parameters of the TSMB algorithm as defined in section
 \protect{\ref{sec2.1}} and total statistics in 1000 update cycles
 ($U_k$) of our runs.}
\label{tab_run}
\begin{center}
\begin{tabular}{|c|c|c|c|c|c|c|c|c|r|}
\hline
run &$\beta$&$\kappa$&$n_1$&$n_2$&$n_3$&$n_4$&$\lambda$&$\epsilon$& 
$U_k$ \\ \hline
$(a)$ &5.28&0.160 &  20&  40&   70& 100  & 2.8 & 1.75 $\cdot 10^{-2}$ &
 80 \\ \hline
$(b)$ &5.04&0.174 &  28&  90&  120& 150  & 3.0 & 3.75 $\cdot 10^{-3}$ & 
 33 \\ \hline
$(c)$ &4.84&0.186 &  38& 190&  240& 300  & 3.6 & 1.44 $\cdot 10^{-3}$ & 
 31 \\ \hline
$(d)$ &4.80&0.188 &  44& 240&  300& 300  & 3.6 & 7.2 $\cdot 10^{-4}$  & 
 12 \\ \hline
$(e)$ &4.76&0.190 &  44& 360&  380& 500  & 3.6 & 2.7 $\cdot 10^{-4}$  & 
144 \\ \hline
$(f)$ &4.80&0.190 &  44& 360&  380& 500  & 3.6 & 2.7 $\cdot 10^{-4}$  & 
224 \\ \hline
$(g)$ &4.72&0.193 &  52& 600&  750& 800  & 3.6 & 0.9 $\cdot 10^{-4}$  & 
196 \\ \hline
$(h)$ &4.68&0.195 &  66& 900& 1200& 1100 & 3.6 & 3.6 $\cdot 10^{-5}$  & 
200 \\ \hline
$(i)$ &4.64&0.197 &  72&1200& 1500& 1400 & 3.6 & 1.8 $\cdot 10^{-5}$  & 
110 \\ \hline
$(j)$ &4.64&0.1975&  72&1200& 1350& 1400 & 4.0 & 2.0 $\cdot 10^{-5}$  & 
  4 \\ \hline
\end{tabular}
\end{center}
\end{table*}

 Most of the runs have been done with 32-bit arithmetics.
 Exceptions are run $(j)$ and about 10\% of the statistics in run $(h)$
 where 64-bit arithmetics was used.
 In general, on the $8^3 \cdot 16$ lattice it is not expected that
 single precision makes any difference.
 In fact, the double precision results in run $(h)$ were compatible
 within errors with the single precision ones.

\subsection{Physical quantities}\label{sec3.1}

 In order to monitor lattice spacing and quark mass one has to determine
 some physical quantities containing the necessary information.
 As discussed before, we define the physical distance scale from the
 value of the Sommer scale parameter $r_0$.
 Once $r_0$ in lattice units is known one can transform any dimensionful
 quantity, for instance the pion mass $m_\pi$, from lattice to
 physical units.
 Therefore a careful determination of $r_0/a$ is important.
 For a dimensionless quark mass parameter one can use $M_r$ as defined
 in (\ref{eq1.03}): $M_r = (r_0/a \cdot am_\pi)^2$.
 In addition, we also measured some other quantities like $f_\pi$,
 $m_\rho$ and another definition of the quark mass $m_q$ for obtaining a
 broader basis for orientation.
 In the next subsections the procedures for extracting these quantities
 will be described in detail.

\subsubsection{Masses and amplitudes}\label{sec3.1.1}

 In order to extract masses and amplitudes we compute the zero-momentum 
 two-point functions depending on the time-slice distance $(x_0-y_0)$
\begin{equation}\label{eq3.01}
C_{XY}(x_0-y_0)=\frac{1}{V_s}\sum_{\bf x,y} \langle X^\dagger(x) Y(y)\rangle\ ,
\end{equation}
 with $x\equiv(x_0,{\bf x})$ and
\begin{eqnarray*}
X(x)&=&Y(x)= P_5(x)\equiv\bar{q^\prime}(x)\gamma_5q(x) \, \ \ \ \ \ \
(C_{PP}(x_0-y_0)), \\ 
X(x)&=&Y(x)=A_0(x)\equiv\bar{q^\prime}(x)\gamma_5\gamma_0q(x) \ \ \ \
(C_{AA}(x_0-y_0)), \\ 
X(x)&=&Y(x)=V_i(x)\equiv\bar{q^\prime}(x)\gamma_5\gamma_iq(x) \,\,\ \ \ \
(C_{V_iV_i}(x_0-y_0)); \\ 
\end{eqnarray*}
 we also consider the mixed correlator with
\[
X(x)=A_0(x) \ ,\ \ \ \ Y(x)=P_5(x)  \ \ \ \ \ \ \ \ \ \ \ \ \, (C_{AP}(x_0-y_0))\ \ .
\]
 Exploiting translation invariance we pick the source $y$ in (\ref{eq3.01}) 
 at random over the lattice.
 Taking into account correlations between different time-slices,
 one sees that this procedure is optimal for the ratio 
 computational cost / final statistical error for hadronic observables.

 Masses and amplitudes are in general obtained from the asymptotic 
 behaviour of the correlators:\footnote{Amplitudes are assumed to be real.}
\begin{eqnarray}\label{eq3.02}
C_{XY}(T)&=&\frac{\xi_{XY}^2}{2 m_p} \, \, 
(e^{-m_p T}+(-1)^{X+Y} e^{-m_p(L_t-T)})\\   
\xi_{XY}&=&\sqrt{\langle 0|X(0)|p\rangle \langle 0|Y(0)|p\rangle} \ ,
\end{eqnarray}
 where $|p\rangle$ is the zero-momentum state of the particle associated
 with the operators $X(x)$ and $Y(x)$, $m_p$ the corresponding mass and 
 $(-1)^{X(Y)}$ the time-parity of $X(Y)(x)$.
 We determine parameters $m_p$ and $\xi_{XY}$ by global fitting
 over a range of time-slice distances  (after time- symmetrization)
 $T\in[T_{min},L_t/2]$. We find the optimal value for $T_{min}$ 
 by checking the behaviour of the effective local mass $m_{eff}(T)$. 
 The latter is implicitly defined by the relation
\begin{eqnarray}\label{eq3.03}
&& \frac{C_{XY}(T)}{C_{XY}(T+1)}= \\ \nonumber
&& \,\,\,\, \frac{e^{-m_{eff}\,(T)
  T}+(-1)^{X+Y}e^{-m_{eff}(T)\,(L_t-T)}}{e^{-m_{eff}\,(T)
  (T+1)}+(-1)^{X+Y}e^{-m_{eff}(T)\,(L_t-T-1)}}\ .
\end{eqnarray}
 The value of $T_{min}$ is fixed by the onset of the plateau for $m_{eff}(T)$ 
 as a function of $T$. The plateau-value for the effective mass 
 is always consistent with the result from the global fit procedure.
 The latter gives however the most precise determination.
 
 A typical problem associated with small quark masses is a delayed 
 asymptotic behaviour for correlators (i.e. a larger $T_{min}$)
 resulting in large errors for the hadronic observables. 
 This problem was solved by applying Jacobi smearing \cite{Allton:JAC}
 on both source and sink. Jacobi smearing was applied in a different context
 \cite{Campos:GLUINO,Farchioni:SUSYWTI} in the same situation of light 
 fermionic degrees of freedom, and it showed to improve the overlap of 
 the hadronic operators with the bound-state. 
 Amplitudes and decay constants have been determined from  correlators 
 with local operators.

 We determine the  pion mass $m_\pi$ from the asymptotic behaviour 
 of the correlator
 $C_{PP}(T)$. From $C_{PP}(T)$ one can also extract the amplitude 
 $g_\pi=\langle 0|P_5(0)|\pi\rangle$ by identifying $g_\pi=\xi_{PP}$.
 The $\rho$ meson  mass $m_{\rho}$ is determined from the asymptotic behaviour 
 of the correlator 
\begin{equation}
C_{VV}(T)=\frac{1}{3}\sum_{i=1}^3\: C_{V_iV_i}(T)\ .
\end{equation}

 For the determination of the pion decay constant
 $f_{\pi}\equiv m_{\pi}^{-1}\,\langle 0|A_0(0)|\pi\rangle$ we apply two
 different methods. In the first, the amplitude $\langle 0|A_0(0)|\pi\rangle$
 is obtained by fitting the asymptotic behaviour of
 the correlator $C_{AA}(T)$ while the pion mass is the one coming from
 $C_{PP}(T)$. In the second method \cite{Baxter:QUEN}, we fit the amplitude-ratio 
\begin{equation}
r_{AP}=\frac{\langle 0|A_0(0)|\pi\rangle}{\langle 0|P_5(0)|\pi\rangle}
\end{equation}
 by using the asymptotic behaviour 
\begin{equation}
\frac{C_{AP}(T)}{C_{PP}(T)}=r_{AP}\tanh[m_{\pi}(L_t/2-T)]
\end{equation}
 where $m_{\pi}$ is fixed at the best-fit value from $C_{PP}(T)$.
 The determination of $f_{\pi}$ is then obtained from the relation
\begin{equation}
f_{\pi}=m_{\pi}^{-1}r_{AP}\,g_\pi
\end{equation}
 using for $g_\pi$ the determination from $C_{PP}(T)$.
 In the region of large and moderate quark masses the second method gives by far the
 most precise determination of $f_{\pi}$. This is generally no more true 
 for very light quarks where data are highly correlated. 
 Here the best determination was picked from the two different methods 
 on a case-by-case basis.

 Using the above determinations we can extract the quark mass defined by  
 the PCAC relation
\begin{equation}
m^{PCAC}_q= \frac{f_{\pi}}{2g_{\pi}}\,m_{\pi}^2\ .
\end{equation}
 The PCAC quark mass gives us a second definition of the physical quark mass 
 alternative to (\ref{eq1.03}) 
\begin{equation}
\mu_r \equiv r_0 m^{PCAC}_q\ .
\end{equation}

 We estimated statistical errors on hadron quantities by applying 
 the Jackknife procedure on blocks of data of increasing size. 
 The same procedure is applied also for the Sommer scale 
 parameter (see next subsection). This method provides us with a definition of the 
 {\em integrated autocorrelation} $\tau_{int}$ of the pion mass.
 Autocorrelations in general will be discussed in section (\ref{sec3.2}).
 The results for the hadronic quantities are listed in table~\ref{tab_resu}.

\subsubsection{Sommer scale parameter}\label{sec3.1.2}

There are several phenomenological models that can be used to get an
estimate for the Sommer scale parameter $r_0$ in nature, and most of 
them point towards a value of $r_0\simeq 0.49 \mbox{fm}$. 
On the lattice $r_0/a$ can be calculated from the static 
quark potential, which is in turn determined from Wilson loops.
The basic idea is simple, but since we want to match all our results to
this parameter it is crucial to get a precise determination. 
To achieve this we follow the method proposed by Michael and
collaborators \cite{Michael:R0,Perantonis:R0} and some details in 
\cite{Allton:R0}.

Using the variational approach of \cite{Luscher:R0} we get matrices $W_{ij}(r,t)$
consisting of $r\times t$ loops of smeared gauge links, where our 
smearing technique of choice is APE-smearing 
(indices $i,j$ label the level of smearing). 
We use for our determinations two and six or two, four and six levels 
of smearing and symmetrize the matrices $W_{ij}$. 
The ratio staple/link is set to $\alpha=0.45$.

From the solutions to
\begin{eqnarray}\nonumber
W_{ij}({\bf r}, t) \phi({\bf r})_j^{(k)}
&=&
\lambda^{(k)}({\bf r}; t, t_0) W_{ij}({\bf r}, t_0)
\phi({\bf r})_j^{(k)},\\
i, j, k &=& 0, 1 (, 2)
\end{eqnarray}
one gets the eigenvector $\phi({\bf r})_j^{(0)}$ for the largest
eigenvalue $\lambda^{(0)}({\bf r}; t=t_0+1, t_0)$. 
This equation is solved by transforming it into an ordinary 
eigenvalue equation, where several ways are possible:
\begin{eqnarray}
&&W({\bf r},t_0)^{-1}W({\bf r},t)\phi = \lambda \phi \label{first}\\
&&W({\bf r},t)W({\bf r},t_0)^{-1} (W({\bf r},t_0)\phi) =
\lambda (W({\bf r},t_0)\phi) \\ \nonumber
&&W({\bf r},t_0)^{-1/2}W({\bf r},t) W({\bf r},t_0)^{-1/2}
(W({\bf r},t_0)^{1/2}\phi) \\ 
&& = \lambda (W({\bf r},t_0)^{1/2}\phi)\ .
\end{eqnarray}
In the literature \cite{Luscher:R0} the third version has been used. However 
this can only be done with extremely good statistics. Otherwise it is 
possible that, due to statistical fluctuations, the matrix $W_{ij}$ gets negative 
eigenvalues making the (real) square root impossible. 
We checked that the first two versions give numerically exactly the same result.
For the final determinations we choose the first version (\ref{first}), 
where one has to be careful about the 
fact that $W({\bf r},t_0)^{-1}W({\bf r},t)$ 
has no longer to be symmetric, complicating the 
calculation of the corresponding eigenvectors.

Once the eigenvector $\phi({\bf r})_j^{(0)}$ has been obtained,
we can project the matrix $W_{ij}$ to the ground state:
\begin{equation}
\tilde W_0({\bf r}, t) = \phi({\bf r})_i^{(0)}
W_{ij}({\bf r}, t) \phi({\bf r})_j^{(0)}\ .
\end{equation}
This correlator leads to good estimates of the ground state energy
\begin{equation}
\tilde E_0({\bf r}, t)=\ln\left(
\frac{\tilde W_0({\bf r}, t)}{\tilde W_0({\bf r}, t+1)}\right)\ .
\end{equation}
The potential V({\bf r}) is estimated
by averaging $E_0({\bf r}, t)/t$ over time extensions $t$ with $t\geq 1$
and weight given by the Jackknife error.
Compared to some other methods this way of extracting the potential seems
to give the most reliable estimates with smallest error bars.

The Sommer scale parameter is defined in terms of the potential as
\begin{equation}
r_0^2\left.\frac{\mbox{d}V}{\mbox{d}r}\right|_{r_0} = 1.65\ .
\end{equation}
Having a reliable static quark potential we can follow \cite{Edwards:R0}
by fitting the potential to
\begin{equation}\label{fit}
V({\bf r}) = V_0 + \sigma r - e [\frac{1}{{\bf r}}]
\end{equation}
with $r = |{\bf r}|$ and $[\frac{1}{{\bf r}}]$ being the tree-level
lattice Coulomb term
\begin{equation}
[\frac{1}{{\bf r}}]=4\pi\int^\pi_{-\pi}\!\!\frac{\mbox{d}^3{\bf
k}}{(2\pi)^3}
\frac{\cos({\bf k}\cdot{\bf r})}
{4\sum_{j=1}^3 \sin^2(k_j/2)}\ .
\end{equation}
Due to the small lattice size we had to drop in (\ref{fit}) 
the additional correction term
$f\cdot\left([\frac{1}{{\bf r}}]-\frac{1}{r}\right)$, which could 
have been used to estimate ${\cal O}(a)$ effects, fixing
$e=\pi/12$. 
Bringing together the above equations we extract $r_0$ from
\begin{equation}
r_0=\sqrt{\frac{1.65-e}{\sigma}}\ .
\end{equation}

\subsection{Autocorrelations}\label{sec3.2}

 The ``cost'' of numerical simulations can be expressed in terms of the
 necessary number of arithmetic operations for obtaining during the
 Monte Carlo update process a new ``independent'' gauge field
 configuration.
 The real cost can be then easily calculated once the price of e.g.
 a floating point operation is known.
 For a definition of the independence of a new configuration the
 {\em integrated autocorrelation} $\tau_{int}$ is used.
 (For a general reference see \cite{Montvay:BOOK}.)
 $\tau_{int}$ does depend on the particular quantity it refers to.
 Of course, it is reasonable to choose an ``important'' quantity as,
 for instance, the pion mass but simple averages characterizing
 the gauge field such as the plaquette average are also often considered.

 In case of the TSMB algorithm a peculiar feature is the reweighting
 step correcting for the imperfection of polynomial approximations.
 As will be discussed in the next subsection, in most of our runs
 this correction is totally negligible but even in these cases it
 is important to perform the reweighting on a small subsample of
 configurations in order to check that the used polynomials are precise
 enough.
 In some cases, especially for very small quark masses, there are
 a few {\em exceptional configurations} with small eigenvalues of
 the squared hermitean fermion matrix ($\tilde{Q}^2$) which are
 practically removed from statistical averages by their small
 reweighting factors.
 In the calculation of expectation values these reweighting factors
 were always taken into account.
 For the autocorrelations the effect of the exceptional configurations
 is in most cases negligible.

\subsubsection{Integrated autocorrelation of the pion mass}\label{sec3.2.0}

 In case of secondary quantities such as the pion mass, or in general
 any function of the primary expectation values, the straightforward
 definition of the integrated autocorrelation $\tau_{int}$ for primary
 quantities is not directly applicable.
 In fact, there are several possibilities which we shall now shortly
 discuss.

 {\em Linearization:} as it has been proposed by the ALPHA Collaboration
 \cite{Frezzotti:BENCHMARK}, in the limit of high enough statistics the
 problem of the error estimate and of the autocorrelation for secondary
 quantities can be reduced to considering a linear combination of primary
 quantities.
 Let us denote the expectation values of a set of primary quantities by
 $A_\alpha,\; (\alpha=1,2,\ldots)$.
 Their estimates obtained from a data sequence are $\overline{a}_\alpha$.
 For high statistics the estimates are already close to the true values:
 $|\overline{a}_\alpha-A_\alpha| \ll 1$.
 Therefore, if the secondary quantity is defined by a function $f(A)$ of
 primary quantities, we have
\begin{equation}\label{eq320.01}
f(\overline{a})-f(A) \simeq \sum_\alpha (\overline{a}_\alpha-A_\alpha)
\frac{\partial f(A)}{\partial A_\alpha} \ .
\end{equation}
 The values of the derivatives are constants therefore on the right hand
 side there is a linear combination of primary quantities which can be
 handled in the same way as the primary quantities themselves.
 Since
\begin{equation}\label{eq320.02}
\frac{\partial f(A)}{\partial A_\alpha} \simeq
\left.\frac{\partial f(A)}{\partial A_\alpha}\right|_{A=\overline{a}}
\equiv \bar{f}_\alpha \ ,
\end{equation}
 one can consider the linear combinations
\begin{equation}\label{eq320.03}
A_{\bar{f}} \equiv \sum_\alpha A_\alpha \bar{f}_\alpha \ ,
\hspace{1em}
\overline{a}_{\bar{f}} \equiv
\sum_\alpha \overline{a}_\alpha \bar{f}_\alpha
\end{equation}
 and the variance of the secondary quantity can be estimated as
\begin{equation}\label{eq320.04}
\sigma_f^2 \simeq
\left\langle (\overline{a}_{\bar{f}}-A_{\bar{f}})^2
\right\rangle \ .
\end{equation}
 (Note that here $\langle\ldots\rangle$ stands for the expectation value
 in an infinite sequence of identical measurements with the same
 statistics as the one under consideration.)
 According to (\ref{eq320.04}) the integrated autocorrelation of the
 secondary quantity can be defined as the integrated autocorrelation
 of $A_{\bar{f}}$.

 This way of obtaining error estimates and autocorrelations of secondary
 quantities is simple and generally applicable.
 Let us note that because of the reweighting even the simplest
 physical quantities are given by ratios of two expectation values and
 are, therefore, secondary quantities.

 {\em Blocking:} in case of sufficiently large statistical samples the
 integrated autocorrelations of secondary quantities can also be obtained
 by comparing statistical fluctuations of data coming from the measurement
 program before and after a blocking procedure.
 The blocking procedure eliminates for increasing block size the
 autocorrelations between data and the final error is the one for
 {\em uncorrelated data}.
 Since the latter is the true error of the measurement, it is appropriate
 to use this definition  of $\tau_{int}$ to estimate  the real cost of a
 simulation.
 In the case of primary quantities such as the plaquette this definition
 coincides with the usual one.

 For a generic quantity $A$ one can define $\sigma^B_n(A)$ as the standard
 deviation of the data at blocking-level  $n$. In the case of the pion mass,
 we determine this quantity by applying the
 jackknife procedure on the hadron correlators averaged over blocks of length
 $n$. In the limit of infinite statistics, for increasing $n$,
 $\sigma^B_n(A)$ should approach after a transient an asymptotic value
 corresponding to the standard deviation of the
 uncorrelated data $\sigma_{unc}(A)$. For finite statistics, $\sigma^{B}_n(A)$
 fluctuates around $\sigma_{unc}(A)$. We determine  $\sigma_{unc}(A)$
 by averaging $\sigma^{B}_n(A)$ over a range of block sizes $n$ after the transient.
 The error on this determination is given by the mean dispersion of data
 around the average.
 Once $\sigma_{unc}(A)$ is given the integrated autocorrelation is defined as
\begin{equation}\label{eq3.04}
\tau_{int}=\frac{1}{2}\left(\frac{\sigma_{unc}(A)}
                                                 {\sigma^B_1(A)}\right)^2 \ .
\end{equation}
 Another way of writing the above formula is
\begin{equation}\label{eq3.05}
 N_{unc}=\frac{N_{stat}}{2\tau_{int}}
\end{equation}
 where $N_{stat}$ is the original statistics and $N_{unc}$ is the number of
 uncorrelated configurations; so $2\tau_{int}$ can be thought of
 as the distance between two uncorrelated configurations.

 {\em Covariance matrix:} in most cases it is a good approximation to assume
 that the probability distribution of the estimates $\overline{a}_\alpha$ of
 the primary quantities $A_\alpha$ is Gaussian:
\begin{equation}\label{eq3.06}
P(\overline{a}) \propto \exp\left\{ -\frac{1}{2}\sum_{\alpha\alpha^\prime}
(\overline{a}_\alpha-A_\alpha) \,C^{-1}_{\alpha\alpha^\prime}\,
(\overline{a}_{\alpha^\prime}-A_{\alpha^\prime}) \right\} \ .
\end{equation}
 The covariance matrix is
\begin{equation}\label{eq3.07}
\langle (\overline{a}_\alpha-A_\alpha)
(\overline{a}_{\alpha^\prime}-A_{\alpha^\prime}) \rangle =
\langle \overline{a}_\alpha \overline{a}_{\alpha^\prime} \rangle -
\langle \overline{a}_\alpha \rangle \langle \overline{a}_{\alpha^\prime}
\rangle = C_{\alpha\alpha^\prime} \ .
\end{equation}
 The elements of the covariance matrix can be estimated from the data
 sequence by determining the integrated autocorrelations
 $\tau_{int}^{(A_\alpha A_{\alpha^\prime})}$:
\begin{equation}\label{eq3.08}
C_{\alpha\alpha^\prime} \simeq
\left( \overline{a_\alpha a_{\alpha^\prime}}
- \overline{a}_\alpha \overline{a}_{\alpha^\prime} \right)
\frac{2\tau_{int}^{(A_\alpha A_{\alpha^\prime})}}{N_{stat}} \ .
\end{equation}
 Once the probability distribution of the estimates $P(\overline{a})$ is
 known one can obtain an error estimate for any function of
 $\overline{a}_\alpha$ by generating a large number of estimates.
 From the error it is also possible to obtain an indirect estimate of
 the integrated autocorrelation from a formula like (\ref{eq3.04}).

 The integrated autocorrelation of the pion mass (and the error of the
 pion mass) can be obtained by any of these three methods and the results
 are generally consistent with each other.
 The method based on linearization is rather robust already at the level
 of statistics we typically have.
 The blocking method becomes easier unstable, especially for moderate
 statistics.
 This is understandable since the statistics in the individual blocks
 is reduced compared to the total sample.
 The method based on the covariance matrix needs sufficient statistics
 in order that the estimate of the covariance matrix be reliable.
 This is usually the case for effective masses derived from intermediate
 distances but -- in our runs -- this method sometimes fails for the
 largest distances.

\subsubsection{Correction factors}\label{sec3.2.1}

\begin{figure*}
\begin{center}
\includegraphics[angle=-90,width=0.8\colw]{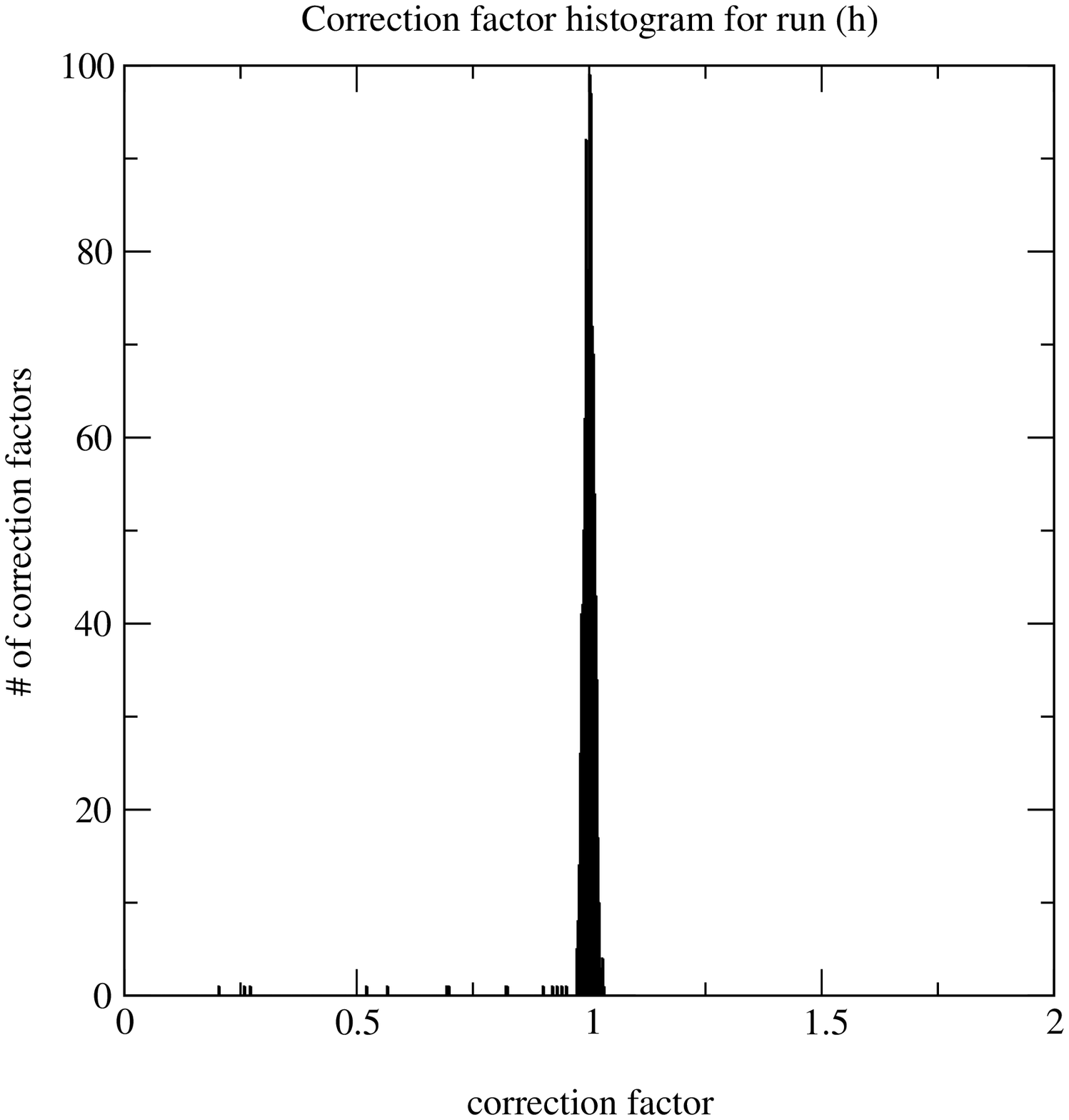}
\hspace{1cm}
\includegraphics[angle=-90,width=0.8\colw]{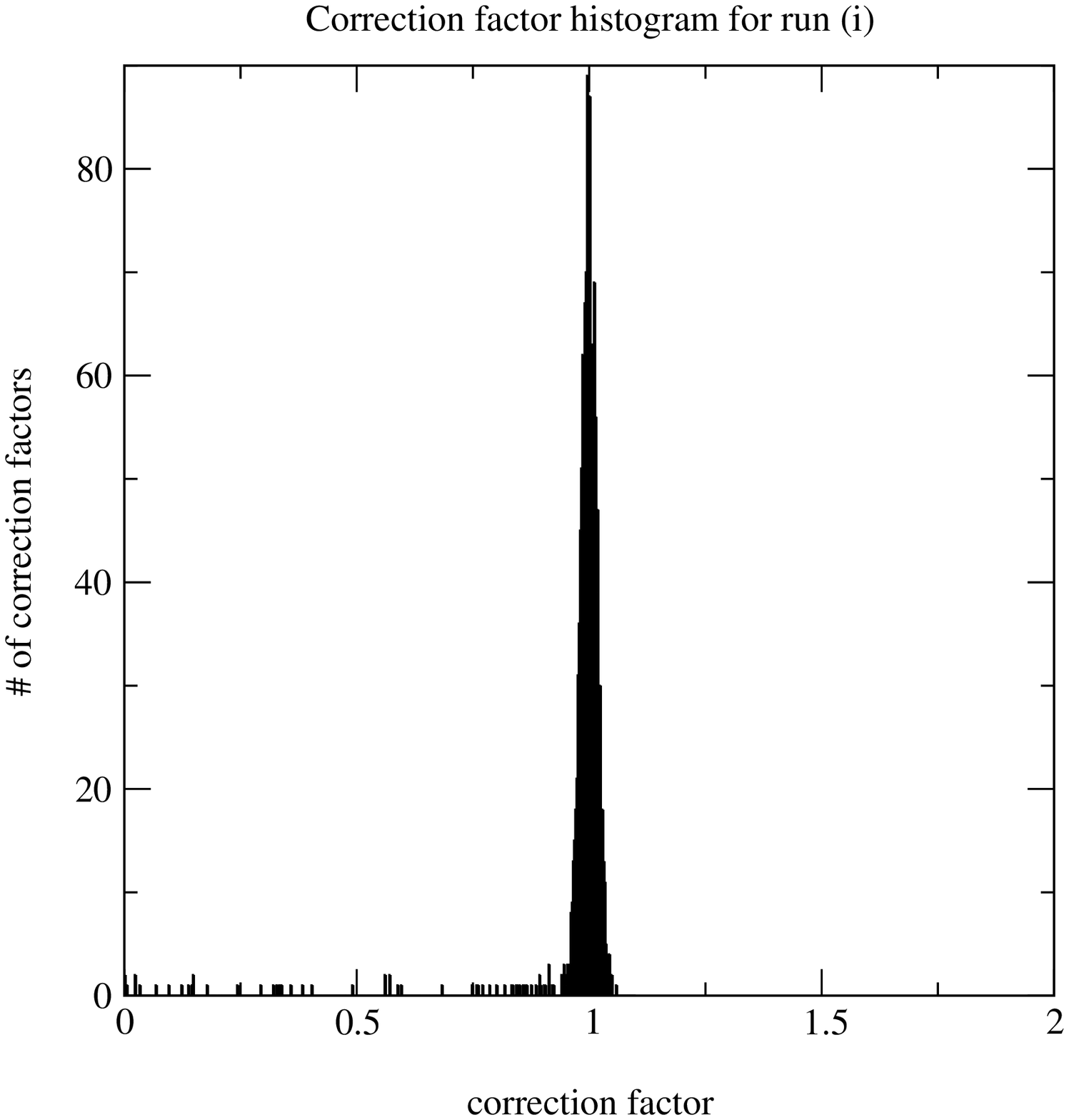}
\end{center}
\vspace{.5cm}
\caption{
 Correction factors for run $(h)$ (left panel) and $(i)$ (right panel). }
\label{fig_cf}
\end{figure*}

\begin{figure}
\begin{center}
\includegraphics[width=\colw,angle=-90]{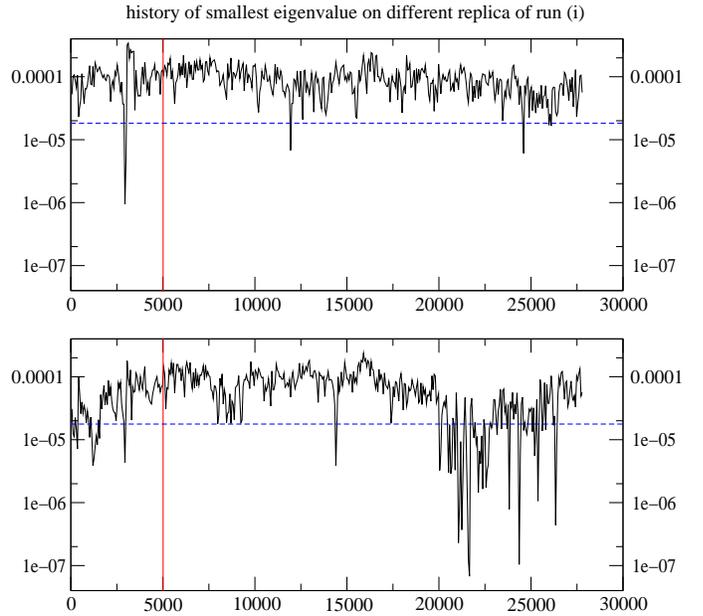}
\end{center}
\vspace{.3cm}
\caption{
 Histories of the smallest eigenvalue at $\beta=4.64$, $\kappa=0.197$
 for two independent lattices.
 The upper figure shows the typical case when the smallest eigenvalue
 stays most of the time above $\epsilon$ shown by the dashed line.
 The lower figure is the history with exceptionally small eigenvalues.
 The measurement of physical quantities was started at the vertical
 line.
}
\label{fig_evs_i}
\end{figure}

 As has been described in subsection \ref{sec2.1} a fourth polynomial
 $P^{(4)}_{n_4}$ can be used to extrapolate to infinite polynomial order
 therefore avoiding the need for several simulations with different
 orders of the second polynomial $P_{n_2}^{(2)}$.
 In our runs the evaluation of the reweighting factors was done in the
 way described in detail in \cite{Campos:GLUINO}.
 Few smallest eigenvalues $\tilde{\lambda}^2$ of the squared hermitean fermion matrix
 $\tilde{Q}^2$, typically four, were
 explicitly determined and the corresponding correction factors were
 exactly taken into account.
 In the subspace orthogonal to the corresponding eigenvectors a
 stochastic estimate based on four Gaussian random vectors was taken.
 (Note that in the limit of infinite statistics a stochastic estimate 
 always gives a correct result independently of the number of random vectors, 
 no systematic error is introduced.)

 As stated before, most of our results were obtained with second
 polynomials $P_{n_2}^{(2)}$ which gave already a good approximation
 of the fermionic measure. In this case the inclusion of the correction 
 factors had nearly no influence on the final determinations
 since they were very close to one.
 Going to smaller quark masses the smallest eigenvalue starts to
 fluctuate more, and it is therefore no longer reasonable to try to use
 a second polynomial that is good enough for all cases.
 Such large fluctuations appeared in runs $(h)$ and $(i)$.
 The histograms of reweighting factors are illustrated by figure
 \ref{fig_cf}.
 It turned out that, as expected, the inclusion of the correction 
 factors had the nice effect of reducing error bars.
 This is especially noticeable for fermionic quantities and 
 in particular the pion mass which is highly correlated
 to the smallest eigenvalue.

 Nevertheless, even in these cases the effect of reweighting was so
 small that the individual estimates with correction factors agreed
 within error bars with those without correction factors.
 As a whole, however, a minor systematic increase in the masses could
 be seen.
 Both histograms in figure \ref{fig_cf} have a
 tail towards small values which are due to eigenvalues that could have
 been further suppressed by a better second polynomial.
 As the figure shows, this tail is more important in run $(i)$ than
 in run $(h)$.
 A closer look at the smallest eigenvalue histories in run $(i)$ reveals
 that the tail near zero was produced by one of the four independent
 parallel lattices when the smallest eigenvalue stayed for some time
 below the lower limit of the approximation interval $\epsilon$ (see
 figure \ref{fig_evs_i}).

 Configurations with small eigenvalues $\tilde{\lambda}^2$
 are interesting because exceptionally small 
 values could indicate crossing of real eigenvalues of the 
 Wilson-Dirac matrix $Q$ to the negative axis. This could give a negative
 sign for the determinant of a single quark flavour.
 We systematically analyzed in all our runs configurations with small 
 $\tilde{\lambda}^2$
 searching for this effect. In the present study 
 we found, for the first time in a QCD simulation with TSMB, 
 configurations with  
 real negative eigenvalues of the Wilson-Dirac matrix. This happened namely in 
 one run, run $(i)$. The (three) configurations are however 
 statistically insignificant, since the corresponding reweighting factors
 are extremely small: 
 $2.7 \cdot 10^{-2},\; 8.2 \cdot 10^{-4}$ and $3.0 \cdot 10^{-4}$.
 Statistically they represent less than 0.03 configurations
 in a sample with a total statistical weight of about 1600.

\subsubsection{Results for autocorrelations}\label{sec3.2.2}

 The analysis of the runs specified in table \ref{tab_run} gives the
 results for physical quantities collected in table \ref{tab_resu}.
 The integrated autocorrelations, where they could be determined, are
 given in table \ref{tab_auto}.

\begin{table*}
\caption{
 Results of runs specified in table \protect{\ref{tab_run}} for
 different physical quantities defined in the text.
 The values given in lattice units can be transformed to physical
 units by canceling the lattice spacing $a$ with the help of the
 results for $r_0/a$ and using $r_0 = 2.53\,{\rm GeV}^{-1}$.}
\label{tab_resu}
\begin{center}
\begin{tabular}{|c|l|l|l|l|l|l|l|}
\hline
\multicolumn{1}{|c|}{$\mbox{run}$} & 
\multicolumn{1}{|c|}{$r_0/a$} & 
\multicolumn{1}{|c|}{$af_\pi$} & 
\multicolumn{1}{|c|}{$am_\pi$} & 
\multicolumn{1}{|c|}{$am_\rho$} & 
\multicolumn{1}{|c|}{$m_\pi/m_\rho$} & 
\multicolumn{1}{|c|}{$M_r$} & 
\multicolumn{1}{|c|}{$\mu_r$}
\\ \hline
$(a)$ &1.885(30)&0.3738(50)&1.2089(36)&1.2982(32)&0.9312(17)&5.19(20)& 
0.498(12)\\ \hline
$(b)$ &1.715(20)&0.4321(23)&1.0428(41)&1.1805(38)&0.8834(14)&3.20(10)& 
0.305(6) \\ \hline
$(c)$ &1.616(110) &0.4171(47)&0.7886(40) &1.0251(48) &0.7693(32)  &1.61(24)& 
0.148(11)\\ \hline
$(d)$ &1.903(159)&0.4199(75)&0.753(11)&0.999(12)&0.752(11)&2.05(40)& 
0.155(13) \\ \hline
$(e)$ &1.697(46)&0.4191(20)&0.7151(20)&0.9941(19) &0.7187(16)& 1.473(88)& 
0.1229(41)\\ \hline
$(f)$ &1.739(33)&0.3658(34)&0.5825(34)&0.9089(47)&0.6431(33)&1.026(51)& 
0.0811(30)\\ \hline
$(g)$ &1.772(41)&0.3791(39)&0.5695(38)&0.9116(33)&0.6256(31)&1.018(61)& 
0.0770(32)\\ \hline
$(h)$ &1.765(37) &0.3668(54)&0.5088(51)&0.8983(35)&0.5675(42)&0.806(50)& 
0.0596(27)\\ \hline
$(i)$ &1.812(46)&0.3575(48)&0.4333(48)&0.8616(80)&0.5002(60)&0.616(45)& 
0.0429(21) \\ \hline
$(j)$ &1.756(128)&0.3377(48)&0.4205(54)&0.859(12)&0.4894(65)&0.545(47)& 
0.0363(38)\\ \hline
\end{tabular}
\end{center}
\end{table*}

 The quoted errors of autocorrelations were estimated in different
 ways. For the determination of the autocorrelation of the pion mass we
 apply the blocking method explained in section \ref{sec3.2.0}.
 In general, one has to say that in some cases our statistics is
 only marginal for a precise determination of the integrated
 autocorrelations. In some cases (run $(a)$, $(j)$) we 
 are not able to quote a reliable result for the autocorrelation of the pion
 mass.
 
 In the high statistics runs with small quark masses $(e)$, $(f)$,
 $(g)$, $(h)$ and $(i)$ we had four independent parallel update
 sequences which could be used for a crude estimate of the errors.
 In addition, whenever the runs were long enough, we used binning with
 increasing bin lengths for the error estimates.

 In general, integrated autocorrelations of the average plaquette are
 longest.
 Those for the smallest eigenvalue are comparable but sometimes by a
 factor 2-3 shorter.
 The important case of $\tau_{int}^{m_\pi}$ is the most favorable
 among the quantities we have considered: it is by a factor 2 to 10
 shorter than $\tau_{int}^{plaq}$.
 Our experience was that the best values for $\tau_{int}^{m_\pi}$
 could be achieved in runs where the lower limit of the approximation
 interval $\epsilon$ was at least by a factor of 2-3 smaller than the
 typical smallest eigenvalue of $\tilde{Q}^2$ and the multi-boson fields
 were relatively often updated by global quasi heatbath.

\begin{figure*}
\begin{center}
\includegraphics[width=0.33\textwidth,angle=-90]{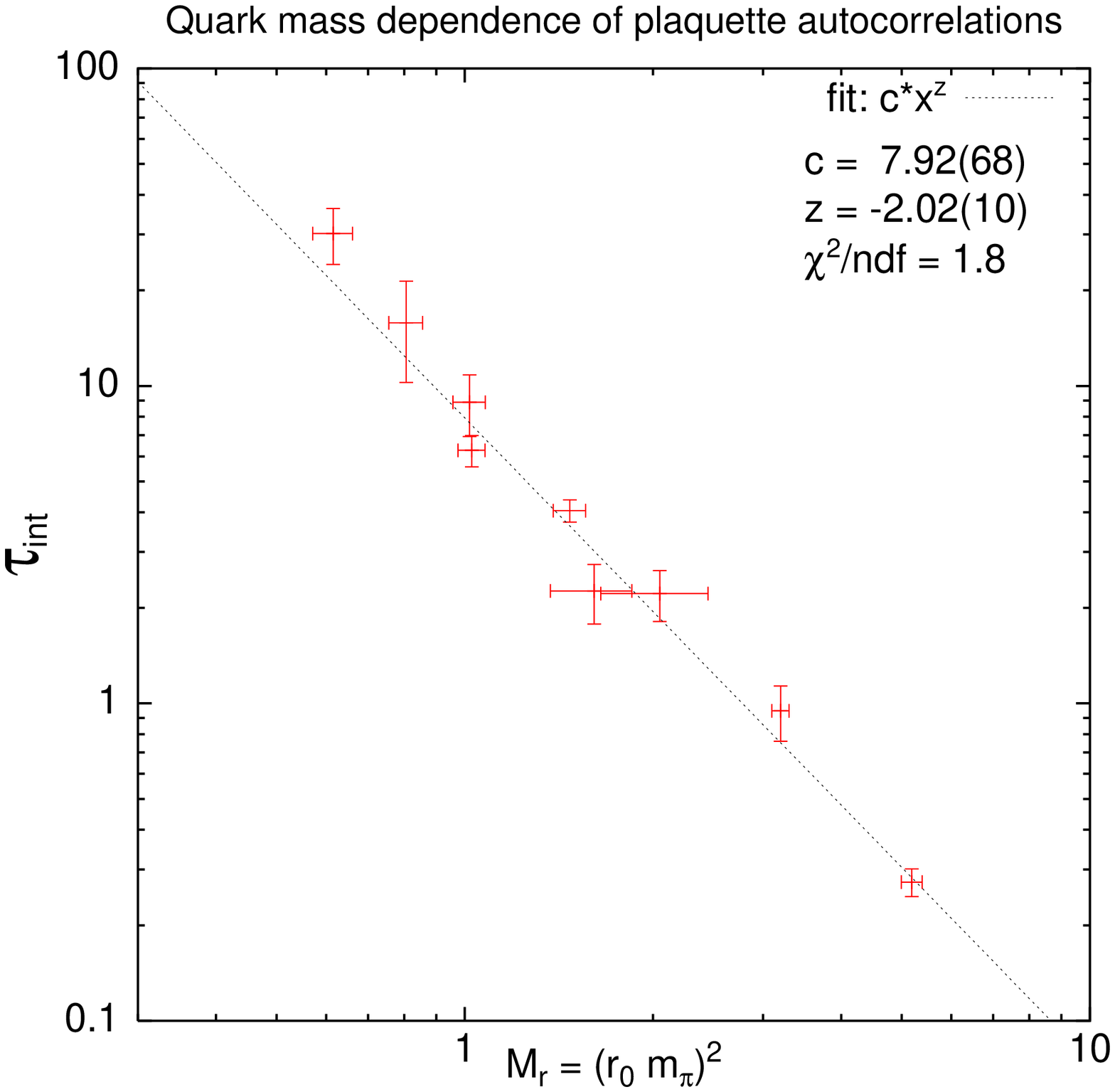}
\includegraphics[width=0.33\textwidth,angle=-90]{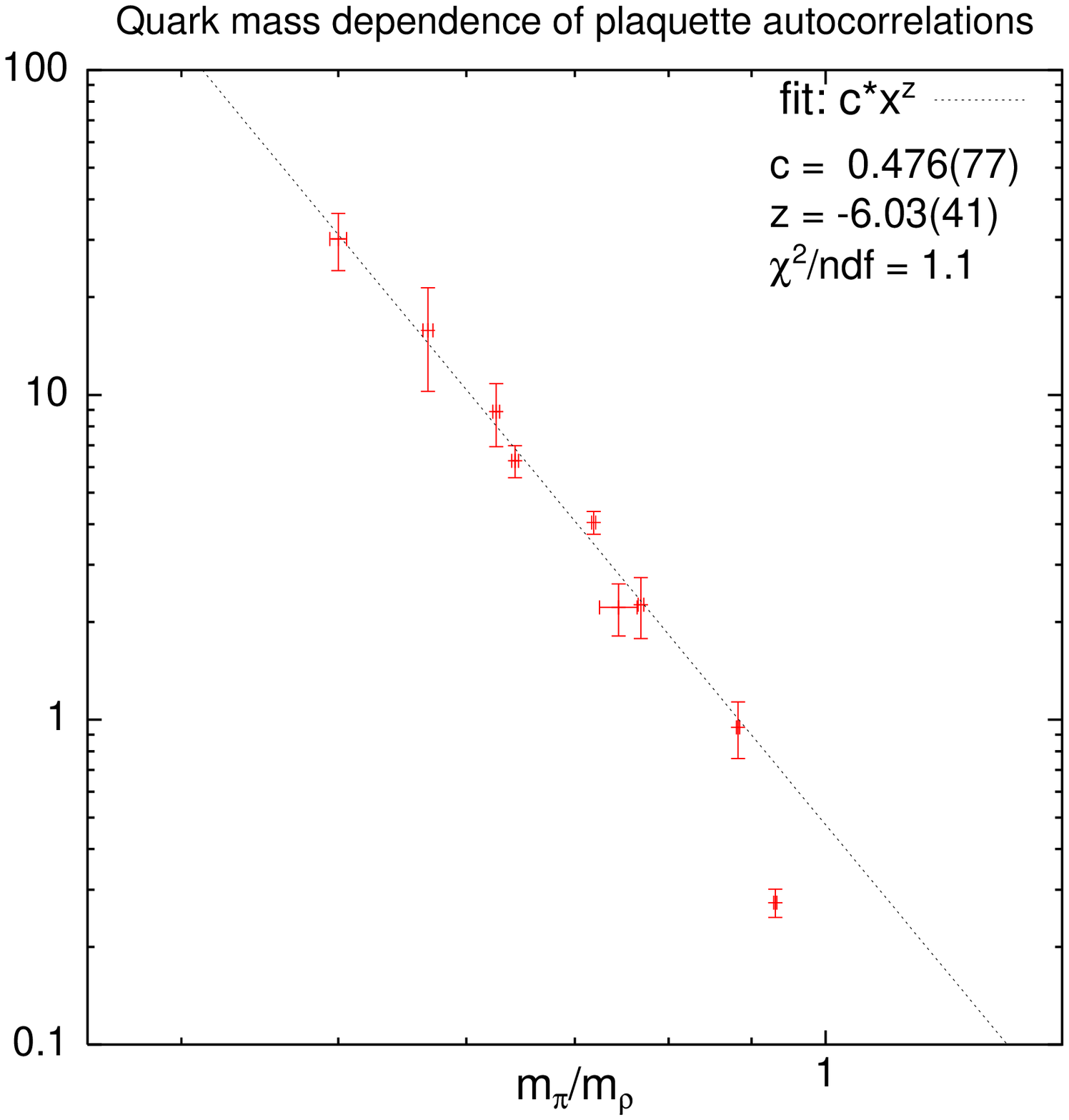}
\includegraphics[width=0.33\textwidth,angle=-90]{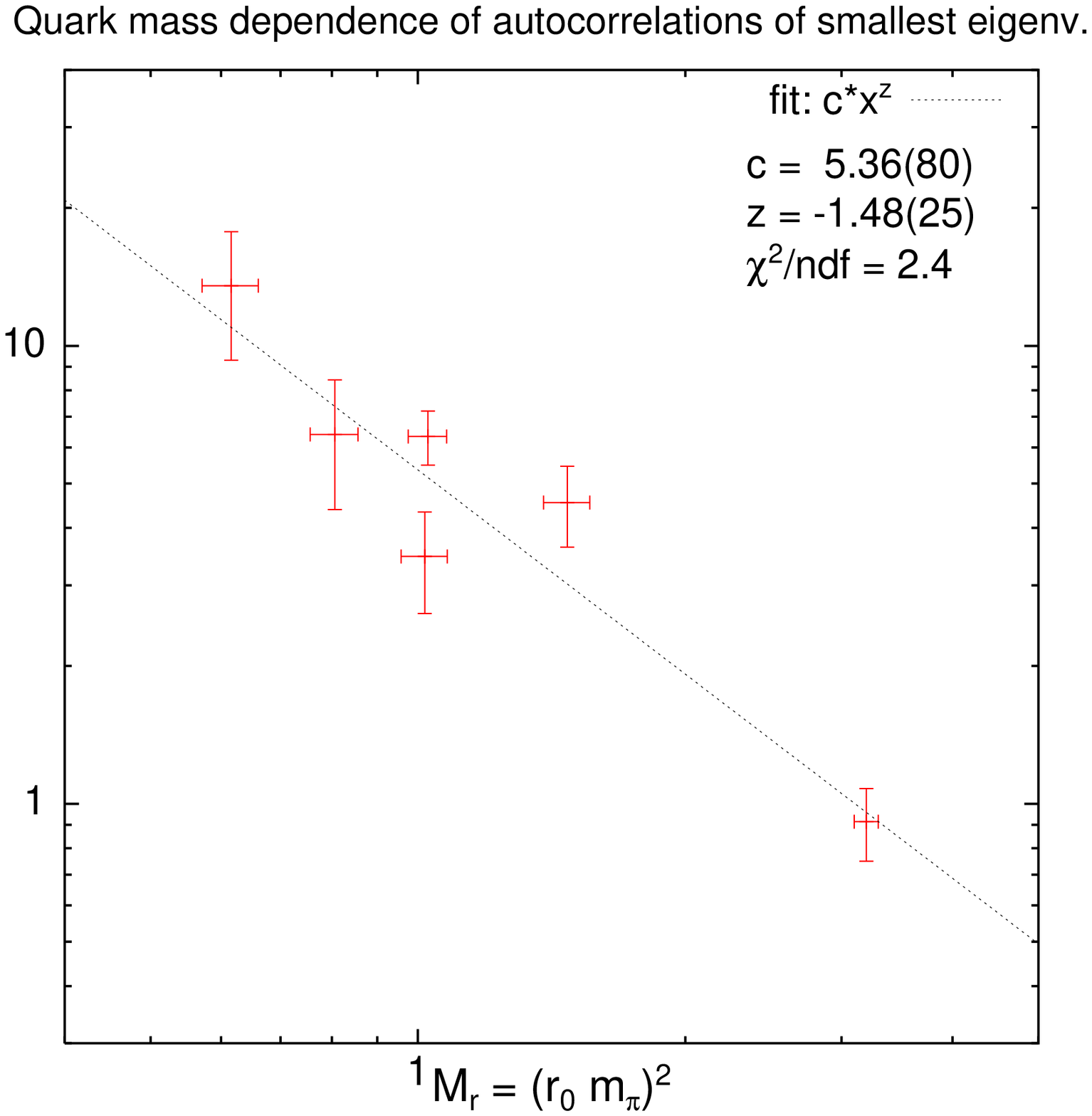}
\end{center}
\caption{
Left panel. 
Power fit of the plaquette autocorrelation given in units of
$10^6 \cdot {\rm MVM}$ as a function of the dimensionless quark mass
parameter $M_r$.
The best fit of the form $c M_r^z$ is at $c=7.92(68),\; z=-2.02(10)$.
Middle panel. 
The same as a function of $m_\pi/m_\rho$.
In this case the last point is omitted from the fit.
The best fit of the form $c (m_\pi/m_\rho)^z$ is at
$c=0.476(77),\; z=-6.03(41)$.
Right panel.
Power fit of autocorrelation for the smallest eigenvalue 
of $\tilde{Q}^2$ given in
units of $10^6 \cdot {\rm MVM}$ as a function of the dimensionless
quark mass parameter $M_r$.
The best fit of the form $c M_r^z$ is at $c=5.36(80),\; z=-1.48(25)$.
}
\label{fig_tau}
\end{figure*}

 Using the values given in table \ref{tab_auto} one can extract, for
 instance, the behaviour of $\tau_{int}^{plaq}$ as a function of the
 dimensionless quark mass parameter $M_r$.
 Since, according to table \ref{tab_resu}, the different runs are
 at slightly different values of $r_0/a$ one can correct the points
 with an assumed power $z_a=2$ to a common value, say, $r_0/a = 1.8$.
 The resulting behaviour is shown by figure \ref{fig_tau} (left panel) where
 a two-parameter fit $c M_r^z$ is also shown.
 The best fit is at $c=7.92(68)$ ($10^6$ MVMs), $z=-2.02(10)$ with a 
 $\chi^2$ per
 number of degrees of freedom of $\chi^2/{\rm d.o.f.}=1.8$.
 (The result for $z$ remains the same if the common value of
 $r_0/a$ is changed in the interval $1.6 \leq r_0/a \leq 2.0$.)

\begin{table}
\caption{
 Integrated autocorrelations in update cycles obtained from runs
 specified by table \protect{\ref{tab_run}}.
 In the second column $C_{cycle}$ gives the number of kMVM's
 ($10^3{\rm MVM}$'s) per update cycle.
 The suffix $min$, $plaq$, $\pi 8$ and $m_\pi$ refer to the minimal
 eigenvalue of $\tilde{Q}^2$, the average plaquette, the pion correlator
 at distance $d=8$ and the pion mass, respectively.}
\label{tab_auto}
\begin{tabular}{|c|c|c|c|c|c|}
\hline
\multicolumn{1}{|c|}{$\mbox{run}$} &
\multicolumn{1}{|c|}{$C_{cycle}$} &
\multicolumn{1}{|c|}{$\tau_{int}^{min}$}   &
\multicolumn{1}{|c|}{$\tau_{int}^{plaq}$}  &
\multicolumn{1}{|c|}{$\tau_{int}^{\pi 8}$} &
\multicolumn{1}{|c|}{$\tau_{int}^{m_\pi}$}
\\ \hline
$(a)$ & 1.49 &              & 200(20)       &          &
\\ \hline
$(b)$ & 2.45 & 340(60)      & 350(50)       & 152(20)  & 140(20)
\\ \hline
$(c)$ & 4.35 &              & 420(80)       &          & 150(20)
\\ \hline
$(d)$ & 5.05 & $\simeq$ 310 & 490(90)       &          & 170(90)
\\ \hline
$(e)$ & 7.34 & 550(110)     & 490(40)       & 274(41)  & 207(33)
\\ \hline
$(f)$ & 7.31 & 810(110)     & 800(90)       & 367(110) & 187(63)
\\ \hline
$(g)$ & 10.5 & 320(80)      & 820(180)      & 466(62)  & 188(13)
\\ \hline
$(h)$ & 16.2 & 380(120)     & 940(330)      & 370(88)  & 186(40)
\\ \hline
$(i)$ & 20.4 & 670(210)     & 1500(300)     & 283(67)  & 153(54)
\\ \hline
$(j)$ & 17.4 & $\simeq$ 390 & $\simeq$ 1050 &          &
\\ \hline
\end{tabular}
\end{table}

 The alternative parametrization in (\ref{eq1.02}) suggests a power fit
 as a function of $m_\pi/m_\rho$.
 A good fit can only be obtained in this case if the last point with
 the largest quark mass is omitted (see figure \ref{fig_tau}, middle panel).
 The best fit parameters are in this case $c=0.476(77),\; z=-6.03(41)$
 with $\chi^2/{\rm d.o.f.}=1.1$.
 The obtained power agrees very well with $z_{\pi\rho}=6$ in
 (\ref{eq1.02}) given by the CP-PACS, JLQCD Collaboration although
 the latter value was obtained in a range of substantially larger 
 quark masses on large lattices.

 The data on the integrated autocorrelation of the smallest
 eigenvalues $\tau_{int}^{min}$ typically have larger errors.
 A fit of the form $c M_r^z$ is shown in figure \ref{fig_tau} (right panel)
 where $c=5.36(80),\; z=-1.48(25)$ with $\chi^2/{\rm d.o.f}=2.4$.
 A fit to the integrated autocorrelation of the pion mass
 $\tau_{int}^{m_\pi}$ gives similar parameters:
 $c=1.99(16),\; z=-1.47(16)$ with $\chi^2/{\rm d.o.f}=1.7$.
 This shows that for $\tau_{int}^{min}$ and $\tau_{int}^{m_\pi}$
 the quark mass dependence is described by $z_\pi \simeq 3$ which is,
 of course, more favorable than $z_\pi \simeq 4$ for $\tau_{int}^{plaq}$.

 Concerning the quality of fits one has to remark that the different
 points belong to individually different optimizations of the
 polynomial parameters which have not necessarily the same quality.
 This implies an additional fluctuation beyond statistics.
 In view of this the $\chi^2$ per number of degrees of freedom values
 are reasonably good.

\section{Eigenvalue spectra}\label{sec4}

 The eigenvalue spectrum of the Wilson-Dirac matrix is interesting
 both physically and from the point of view of simulation algorithms.
 From the physical point of view the low-lying eigenvalues are
 expected to dominate the hadron correlators
 \cite{Ivanenko:LOWEIGEN,Neff:LOWEIGEN} and carry information about
 the topological content of the background gauge field
 \cite{DeGrand:LOWEIGEN,Kovacs:LOWEIGEN,Horsley:LOWEIGEN}.
 Although, as already stressed, in the present work we consider rather
 coarse lattices, given the importance of the question, it is
 interesting to see the effect of the determinant of light quarks
 on the qualitative properties of the eigenvalue spectrum.
 From the algorithmic point of view the knowledge of low-lying
 eigenvalues is crucial for the optimization of polynomial
 approximations.
 Finally, since we plan to perform simulations with an odd number of
 flavors \cite{Farchioni:QQQ}, we have to consider the possibility of
 negative (real) eigenvalues of the non-hermitean quark matrix $Q$,
 which would imply a negative determinant for a single quark flavour.
 For $N_f=2$ the square of the determinant is relevant therefore the
 sign does not matter but the absence (or statistical insignificance)
 of negative eigenvalues at very small quark masses would strongly
 support the assumption that for the heavier strange quark there will
 be no problem with the determinant sign.

 In order to study the low-lying spectrum of the eigenvalues we used
 two methods: for the eigenvalues of the hermitean fermion matrix
 with small absolute value the one by Kalkreuther and Simma
 \cite{Kalkreuter:SIMMA} and for the small eigenvalues of the
 non-hermitean matrix the Arnoldi method
 \cite{Lehoucq:ARPACK,Maschho:PARPACK}.
 The determination of the eigenvalues of the hermitean matrix is in
 general much faster.
 However, the spectrum of the non-hermitean fermion matrix contains
 more informations.
 First of all, the eigenvalues of $Q$ depend trivially on the valence 
 hopping parameter $\kappa_{val}$ because
\begin{equation} \label{eq4.01}
Q = 1 - \kappa_{val} D \ .
\end{equation}
 This is not true for $\tilde{Q}$.
 Moreover, because of the symmetries
\begin{equation} \label{eq4.02}
Q^\dag= \gamma_5 Q \gamma_5 \ , \hspace*{4em} ODO=-D \ , 
\end{equation}
 where $O_{xy}=(-1)^{(x1+x2+x3+x4)} \delta_{xy}$, the spectrum of
 $D$ is invariant under complex conjugation and sign change.
 As a consequence, it is sufficient to compute the low-lying spectrum
 of $Q$ at an arbitrary value $\kappa_{val} = \bar{\kappa}_{val}$.
 Other $\kappa_{val}$ are easily obtained by a shift.
 The value of $\bar{\kappa}_{val}$ is chosen such that it gives the best
 compromise of computation time and precision.

 It turned out that the application of the Arnoldi algorithm is more
 efficient on the even-odd preconditioned matrix $\bar{Q}$ than on $Q$
 itself.
 The analytic relation between the eigenvalues of $\bar{Q}$ and $Q$
 can be used to transform the result back to $Q$.
 Indeed if $Q$ is written in the form
\begin{equation} \label{eq4.03}
Q=1-\kappa \, \left( 
\begin{array}{c c}  0 & D_{eo} \\ D_{oe}& 0 
\end{array}\right)
\end{equation}
 then $\bar{Q}$ is given by
\begin{equation} \label{eq4.04}
\bar{Q}=1-\kappa^2 \, \left( 
\begin{array}{c c}  0 & 0 \\ 0 & D_{oe}D_{eo} 
\end{array}\right) \ .
\end{equation}
 If $v=(v_e,v_o)$ is an eigenvector of $Q$ with eigenvalues $\lambda$
 then it satisfies
\begin{equation} \label{eq4.05}
(\lambda v_e, \lambda v_o ) = 
(v_e - \kappa D_{eo} v_o, v_o - \kappa D_{oe} v_e)
\end{equation}
 and hence
\begin{equation} \label{eq4.06}
(1-\kappa^2 D_{oe}D_{eo}) v_o =
 v_o - (1-\lambda)^2 v_o = \lambda (2-\lambda) v_o \ .
\end{equation}
 As a result, the eigenvalues of $\bar{Q}$ are either 1 (in the even
 subspace) or they satisfy
\begin{equation} \label{eq4.07}
\bar{\lambda}= \lambda (2-\lambda) \ .
\end{equation}
 Because of the symmetries mentioned above, the solutions of
 (\ref{eq4.07}) will give precisely all the eigenvalues of the matrix
 $Q$.
 This relation also gives a possibility to perform a non-trivial check
 of the Arnoldi code.
 (In addition, we also compared the algorithm with a direct
 computation of all the eigenvalues on a small $4^4$ lattice by means of
 a NAG library routine.)
 All checks confirmed the high precision given as an output by the
 ARPACK code, which was, in our cases, always better than  $10^{-4}$
 (relative precision).

\subsection{Small eigenvalues}\label{sec4.1}

 As a first task we computed the low-lying eigenvalues from sample sets
 of 10 configurations for runs in decreasing order of quark masses,
 namely those labeled with $(a)$ and $(c)$-$(j)$ in table \ref{tab_run}.
 In order to have a better access to the most interesting regions of the
 spectrum we analyzed each configuration from two different points of 
 view.
 For each configuration we first determined the 150 eigenvalues of
 the preconditioned Wilson-Dirac matrix ($\bar{Q}$) with
 smallest modulus and then the 50 eigenvalues of the non-preconditioned
 one  ($Q$) with smallest real part.

\begin{figure*}
\begin{center}
\includegraphics[height=0.3\textwidth,width=0.3\textwidth]{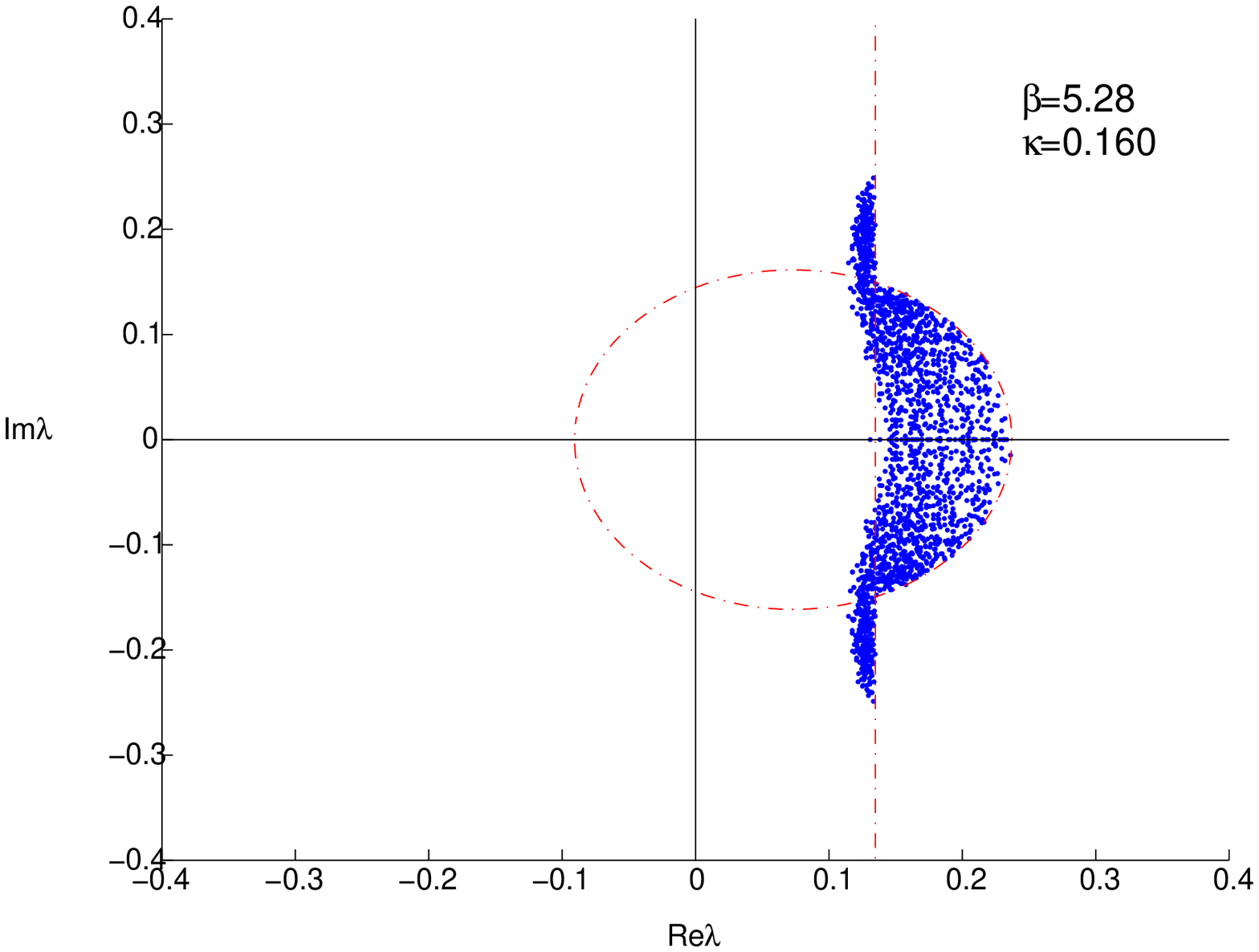}
\includegraphics[height=0.3\textwidth,width=0.3\textwidth]{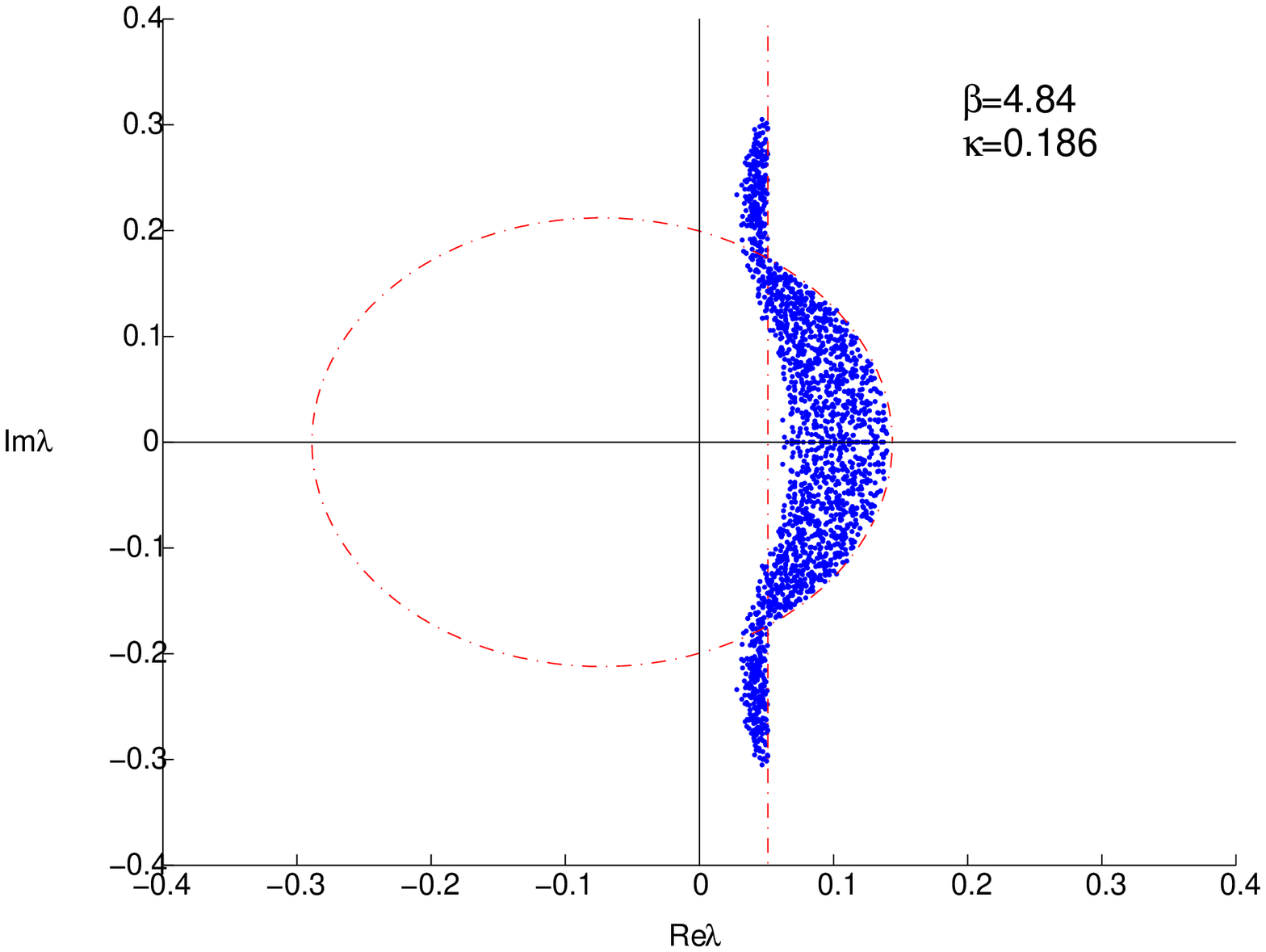}
\includegraphics[height=0.3\textwidth,width=0.3\textwidth]{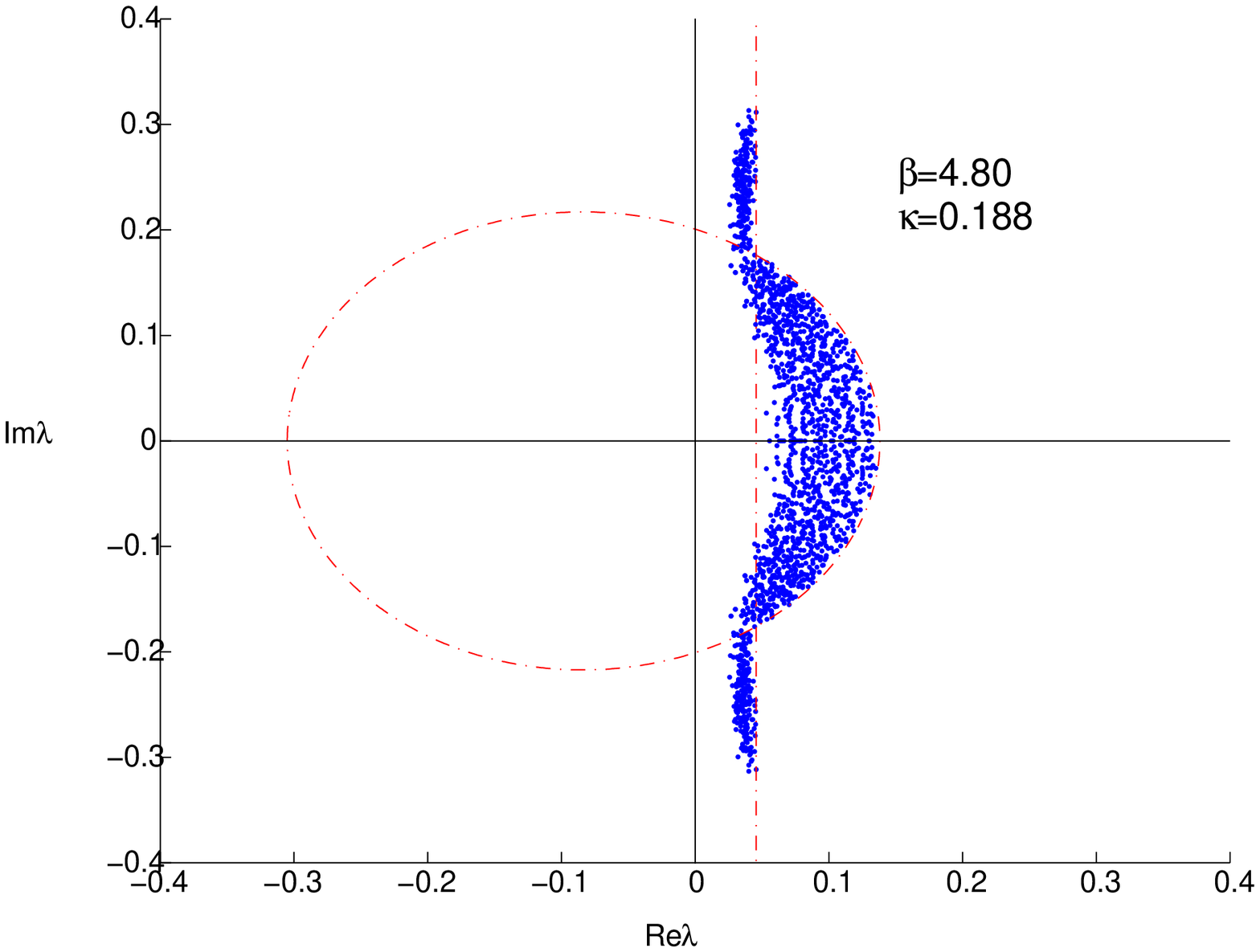}
\includegraphics[height=0.3\textwidth,width=0.3\textwidth]{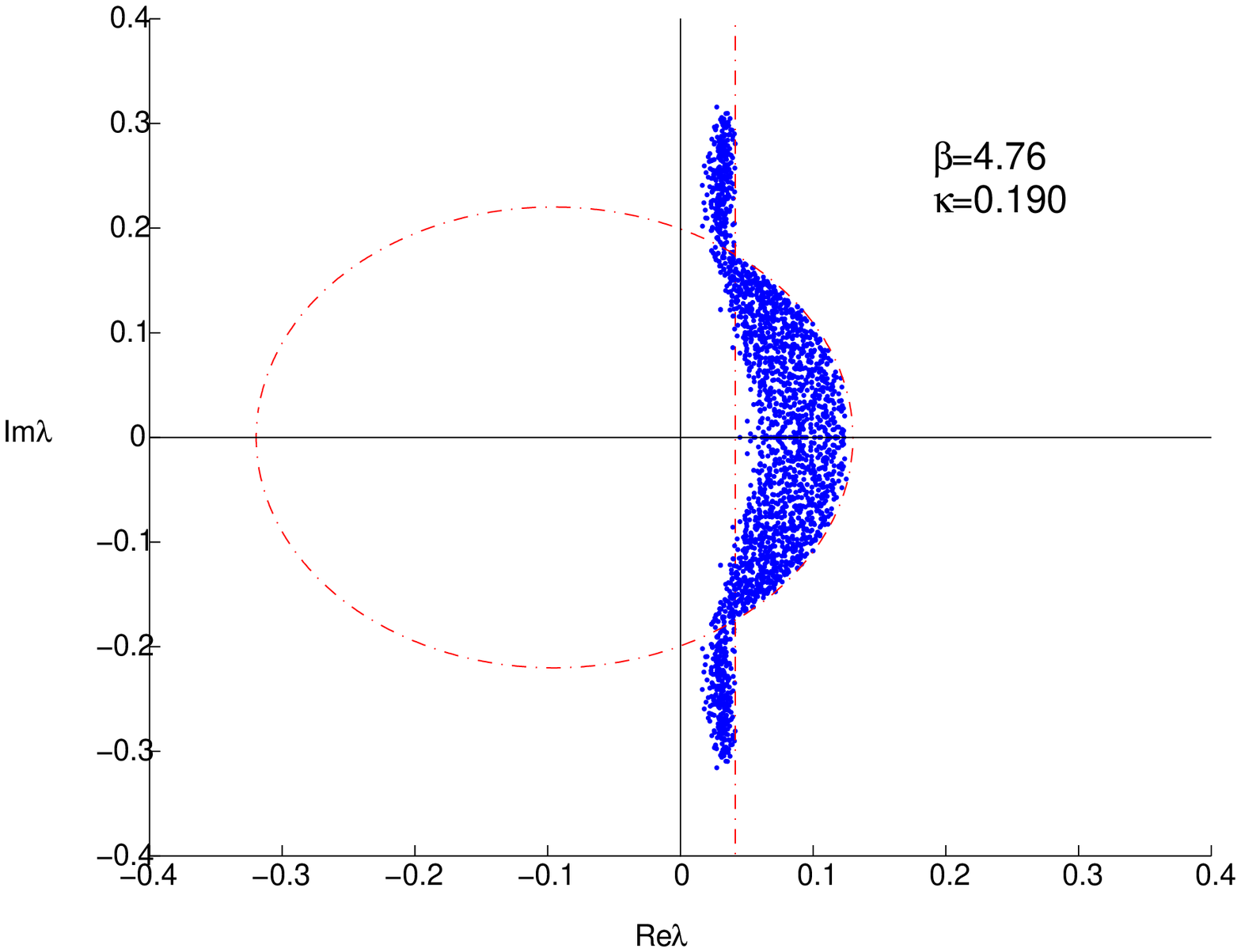}
\includegraphics[height=0.3\textwidth,width=0.3\textwidth]{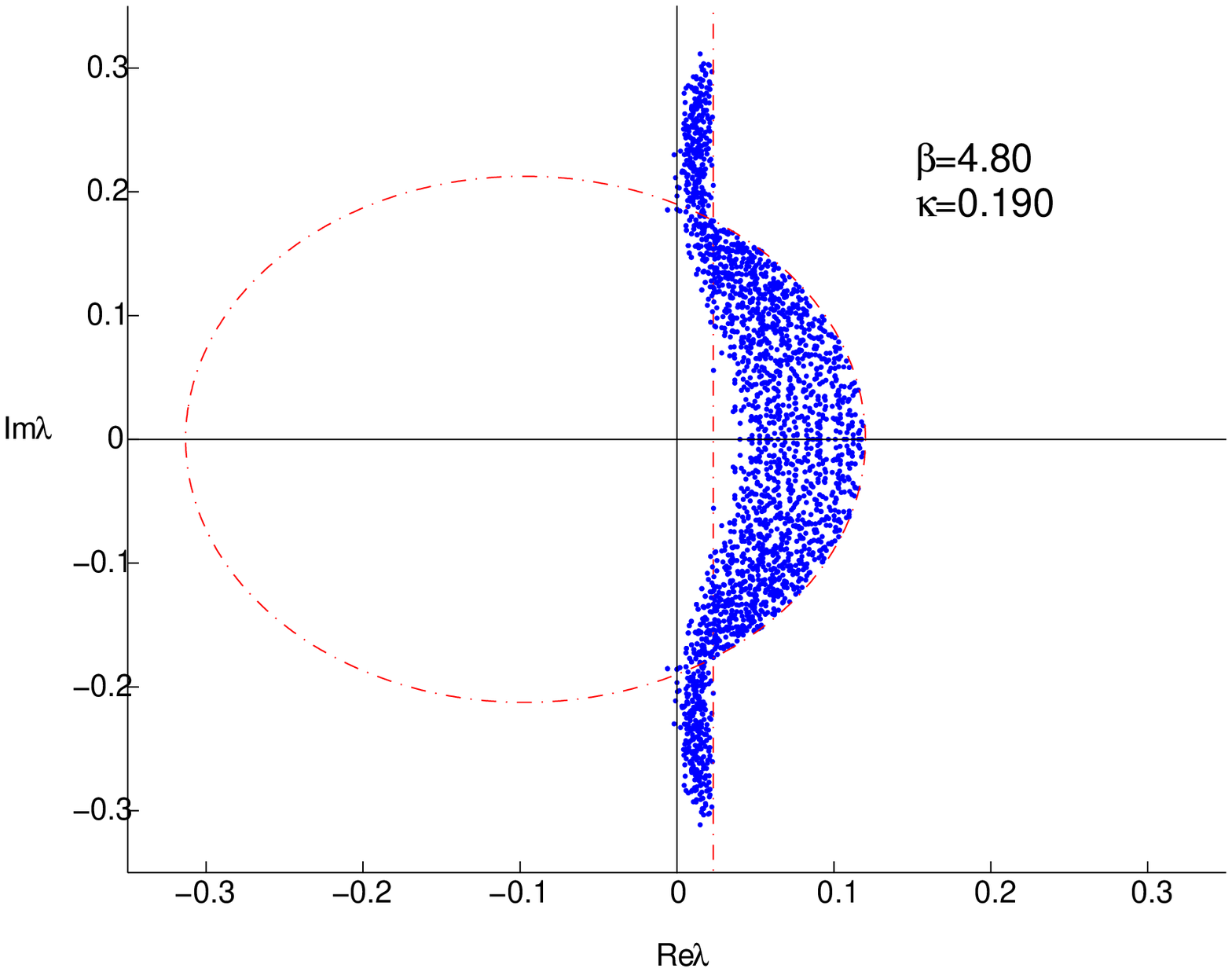}
\includegraphics[height=0.3\textwidth,width=0.3\textwidth]{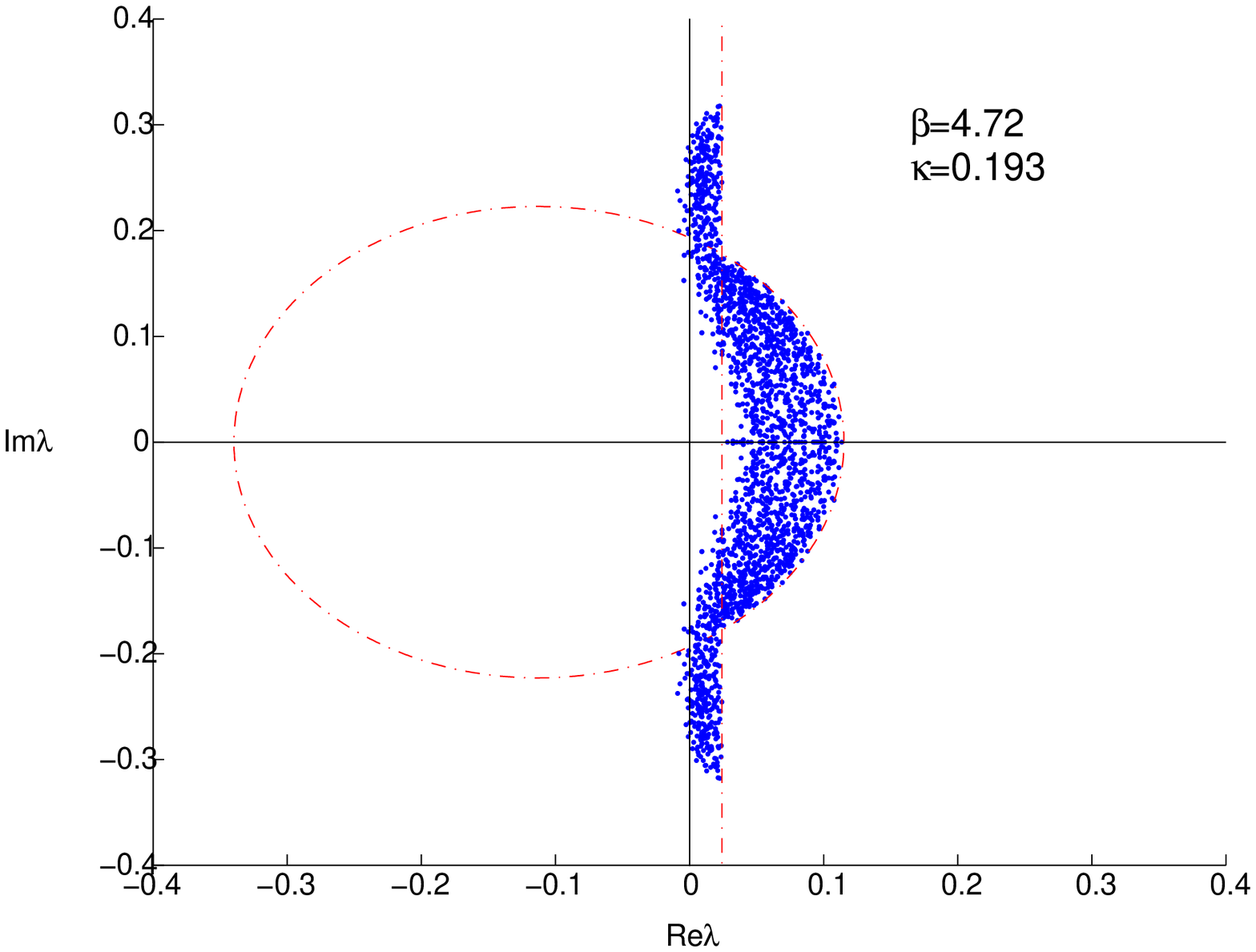}
\includegraphics[height=0.3\textwidth,width=0.3\textwidth]{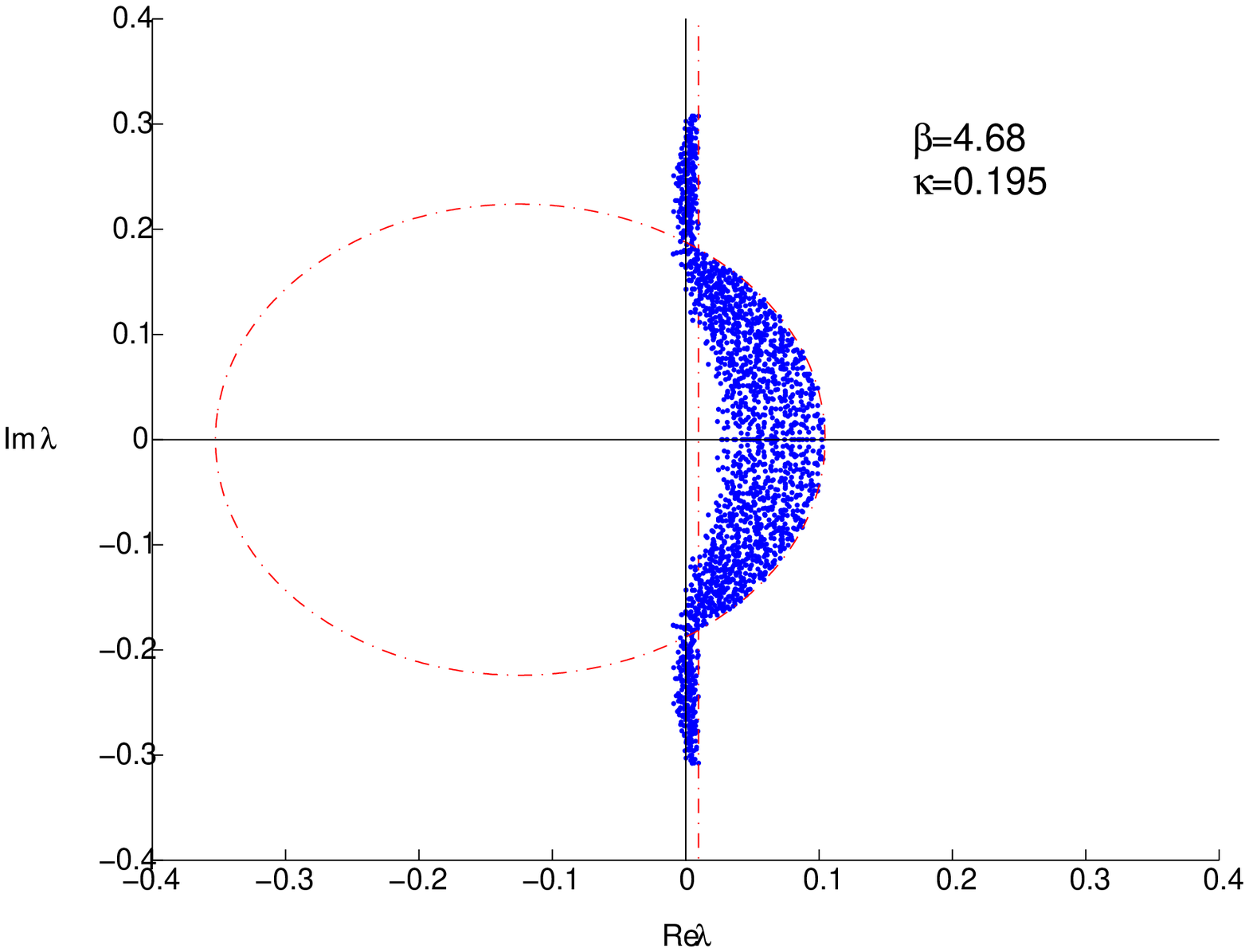}
\includegraphics[height=0.3\textwidth,width=0.3\textwidth]{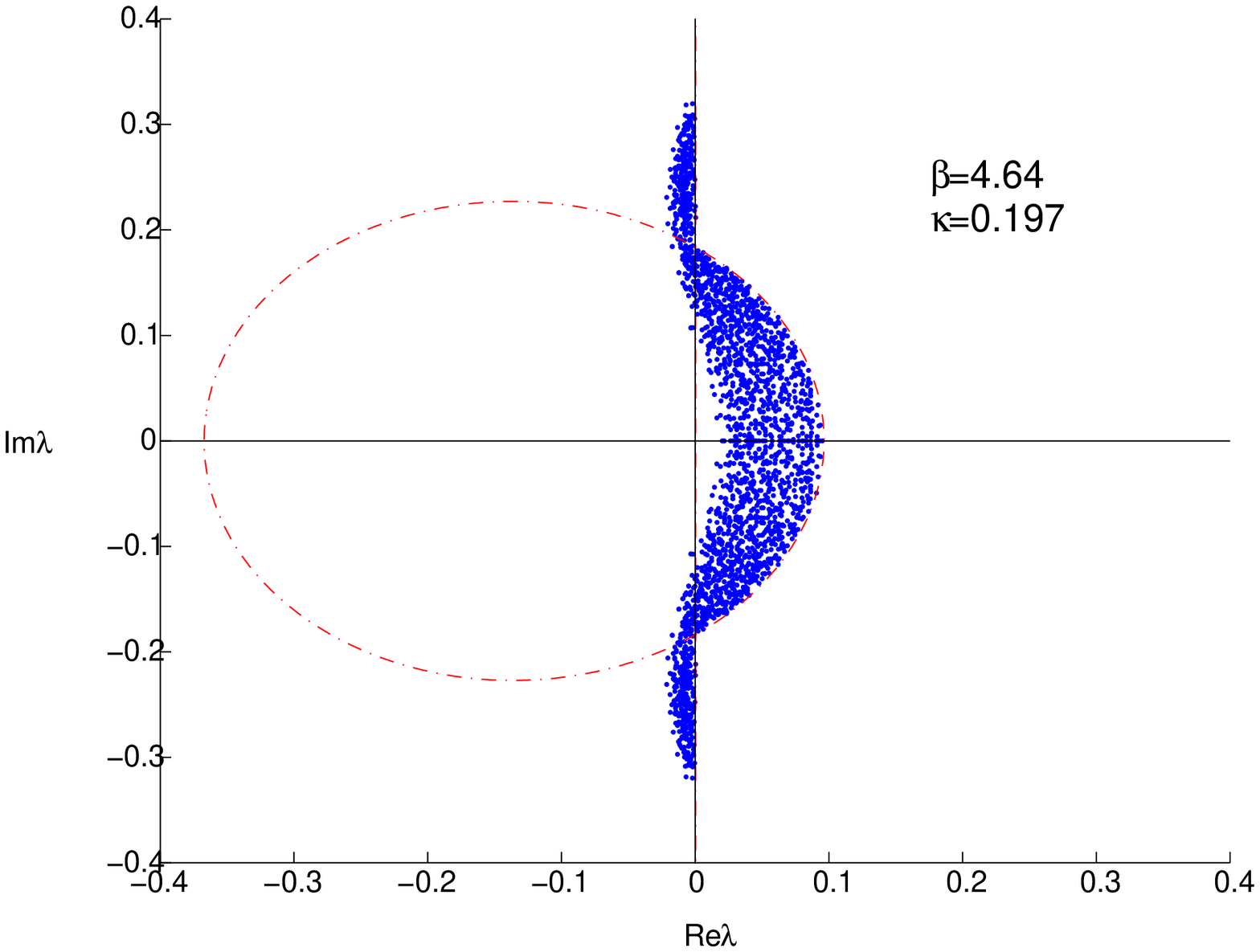}
\includegraphics[height=0.3\textwidth,width=0.3\textwidth]{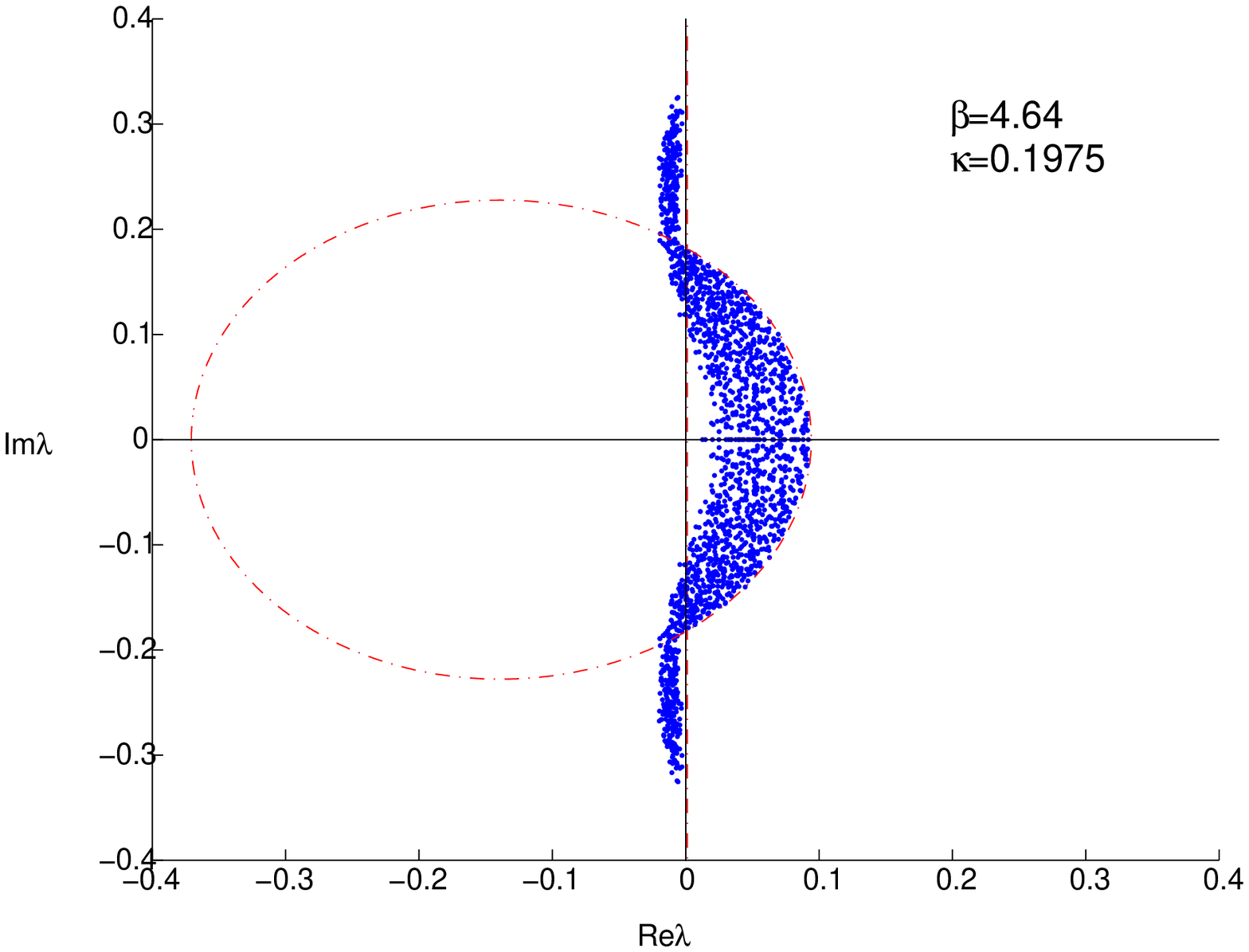}
\end{center}
\caption{
Low-lying eigenvalues from a set of ${\cal O}(10)$ configurations  
for runs $(a)$, $(c)$ to $(j)$.
}
\label{sec4:fig:sys:atoj}
\end{figure*}

 Both computations were performed at an auxiliary value of
 $\kappa_{val}=0.170$, where the Arnoldi algorithm performed better.
 By using the analytical relations (\ref{eq4.01}) and (\ref{eq4.07}) we
 transformed the eigenvalues to those of $Q$ at the $\kappa$ value of
 the dynamical updates ($\kappa\equiv\kappa_{sea}$).
 The results are plotted in figure
 \ref{sec4:fig:sys:atoj}.
 The dashed vertical line shows the limit for the computation of the
 eigenvalues with smallest real part: only the part of the spectrum
 to the left of this line is known.
 In a similar way, by computing the eigenvalues with smallest modulus,
 we have access to the portion of the spectrum inside the dashed circle.
 The circle is deformed and not centered at the origin because it has
 been transformed together with the eigenvalues by using (\ref{eq4.07}).
 In summary, the spectrum is not known in those points of the complex
 plane which are both to the right of the vertical line {\em and}
 outside the circle.

 Since the sequence from $(a)$ to $(j)$ corresponds to decreasing quark
 masses it is not a surprise that the eigenvalues have an increasing
 tendency to go to the left in the complex plane.
 At the same time they are pushed away from zero as an effect of
 including the fermionic determinant in the path integral measure.
 At very small quark masses a pronounced hole near zero is developing.
 For the continuum Dirac operator the spectrum is on a vertical
 line with some gap near zero.
 On our coarse lattices there is an additional horizontal spread of the
 eigenvalues and the picture is strongly deformed.

 The size of the holes produced by the determinant is very important if
 we have in mind the possibility of computing observables at a
 {\em partially quenched} $\kappa_{val}$ higher than $\kappa_{sea}$ used
 in the update.
 The distance between the origin and the smallest real eigenvalue
 determines how much smaller masses (larger $\kappa_{val}$) one can reach by
 partial quenching before encountering exceptional configurations.
 This question can be answered by studying configurations with
 exceptionally small eigenvalues.
 Of course, the reweighting factors discussed in section \ref{sec3.2.1}
 have to be taken into account in this analysis because they suppress
 such configurations to a large extent.

\subsection{Negative eigenvalues}\label{sec4.2}

 One of the purposes of the analysis of the eigenvalues was to determine
 whether there is a statistically significant presence of configurations
 with negative determinant.
 As already said, the sign of the determinant is easy to determine from
 the low-lying spectrum since it is negative if an odd number of real
 negative eigenvalues occurs.
 In fact, non-real eigenvalues always appear in conjugate pairs.
 In the randomly chosen set of configurations reported above we did not
 find a single real negative eigenvalue.
 However, a set of ${\cal O}(10)$ configurations is a rather small
 subsample.

 Additional information on the presence or absence of negative
 eigenvalues in our samples is given by the distribution of reweighting
 factors. Crossing of eigenvalues to negative real axis
 implies small reweighting factors corresponding to very small
 eigenvalues of $\tilde{Q}^2$ below the lower bound
 of the interval $[\epsilon,\lambda]$.
 The calculation of reweighting factors, which was carried out on every
 configuration in the selected subsamples, is much cheaper than the analysis
 of small eigenvalues of the non-hermitean matrix $Q$.
 As we discussed in section \ref{sec3.2.1}, the distribution of
 reweighting factors is strongly peaked near one in all runs, except
 for runs $(h)$ and $(i)$ which have high statistics at very small quark
 masses (see figure \ref{fig_cf}).
 In these cases there are a few configurations with reweighting factors
 close to zero.
 In order to see whether the small reweighting factors (and the
 corresponding small eigenvalues of $\tilde{Q}^2$) are associated to
 negative eigenvalues or not, we concentrated on configurations with
 particularly small eigenvalues of $\tilde{Q}^2$.

 Note that there is no simple analytical relation between the lowest
 eigenvalues of the hermitean and the non-hermitean matrix but it is
 reasonable to expect that small eigenvalues occur together.
 This expectation was confirmed in all cases we investigated.
 An interesting observation was that very small eigenvalues of the
 hermitean matrix seem to be usually associated to small {\em real}
 eigenvalues of the non-hermitean one.
 This is compatible with the fact that real eigenvalues do not need to
 be double degenerate and therefore they can afford to approach closer
 to the origin than a complex conjugate pair.

\begin{figure*}
\begin{center}
\includegraphics[height=0.3\textwidth,width=0.3\textwidth]{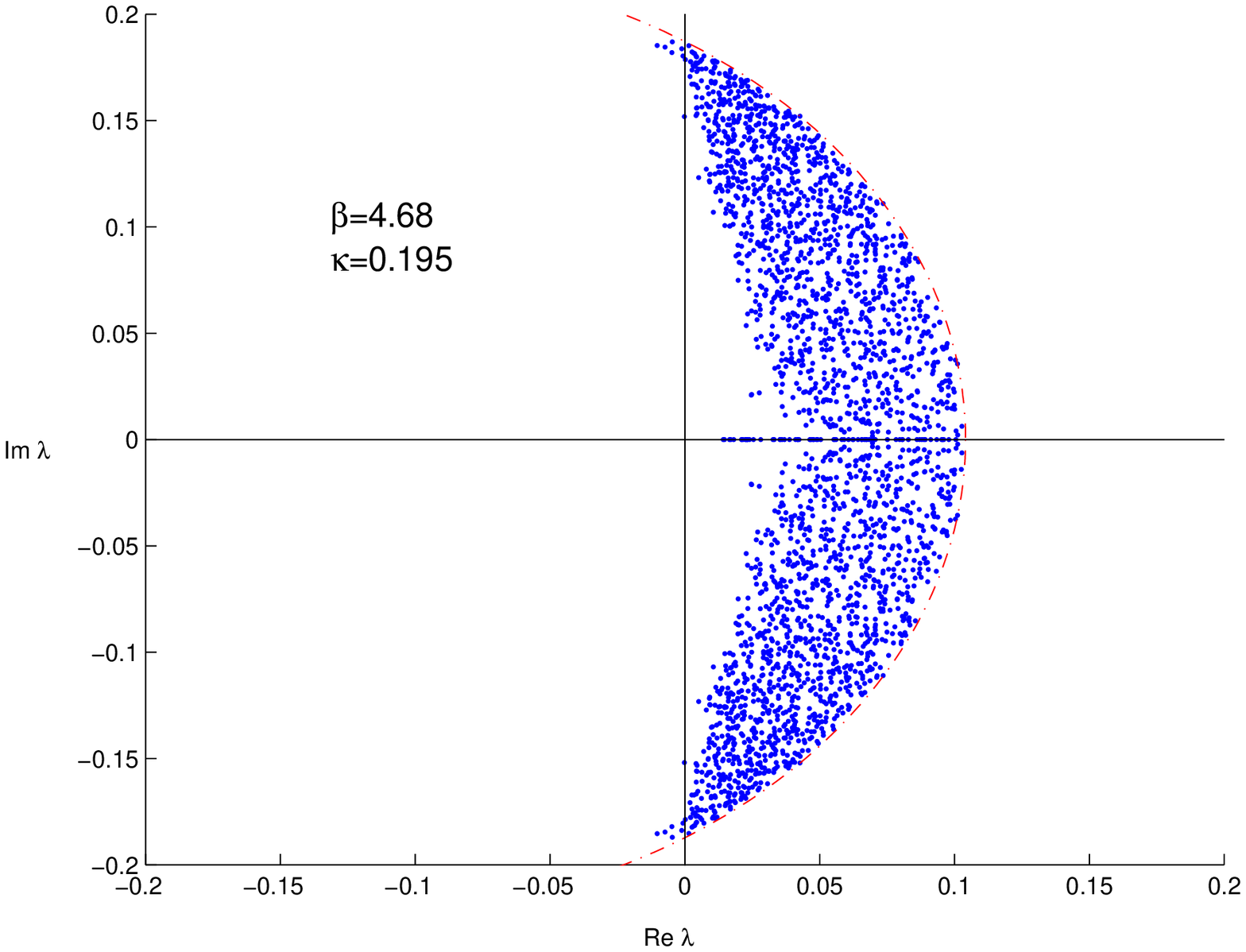}
\includegraphics[height=0.3\textwidth,width=0.3\textwidth]{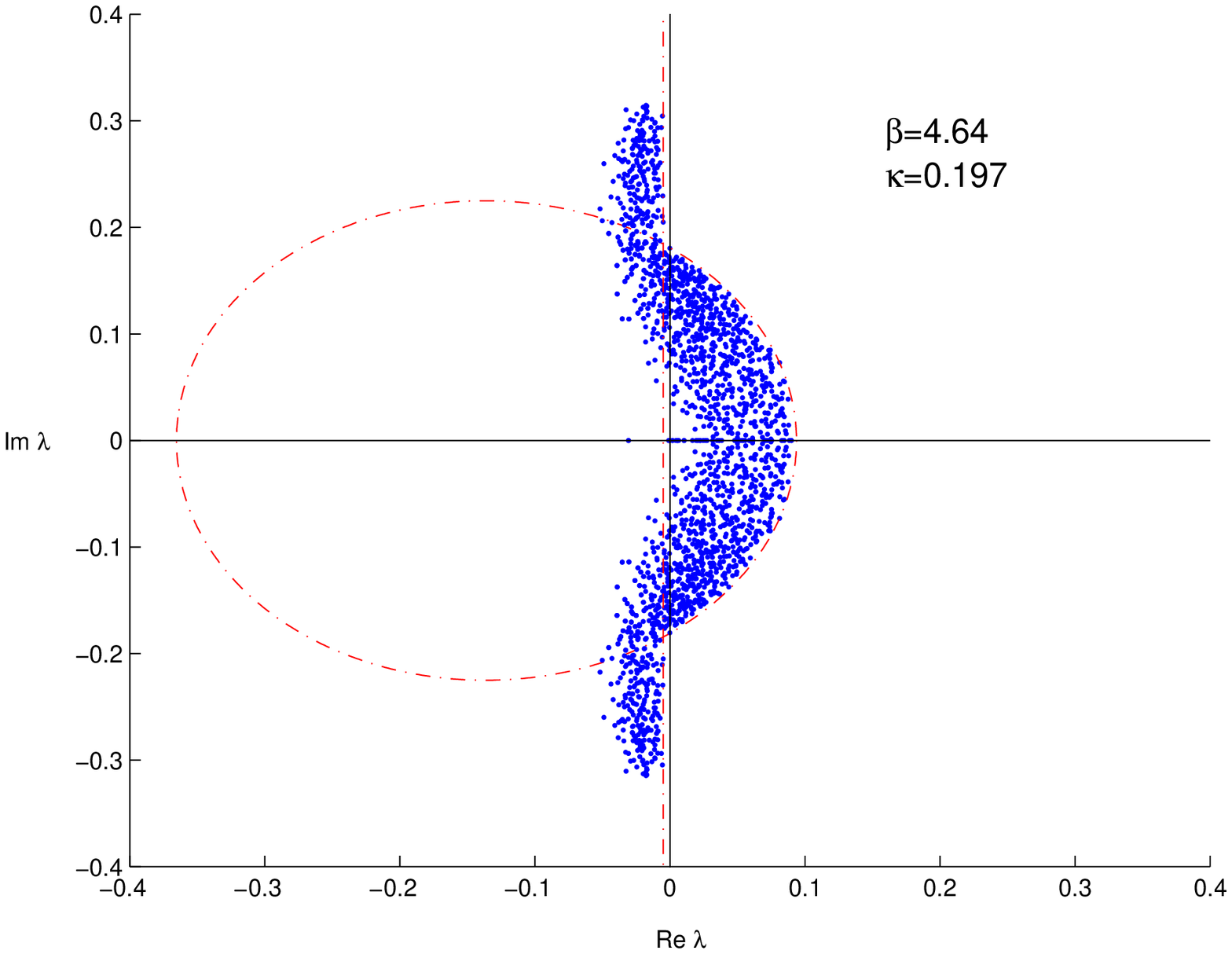}
\includegraphics[height=0.3\textwidth,width=0.3\textwidth]{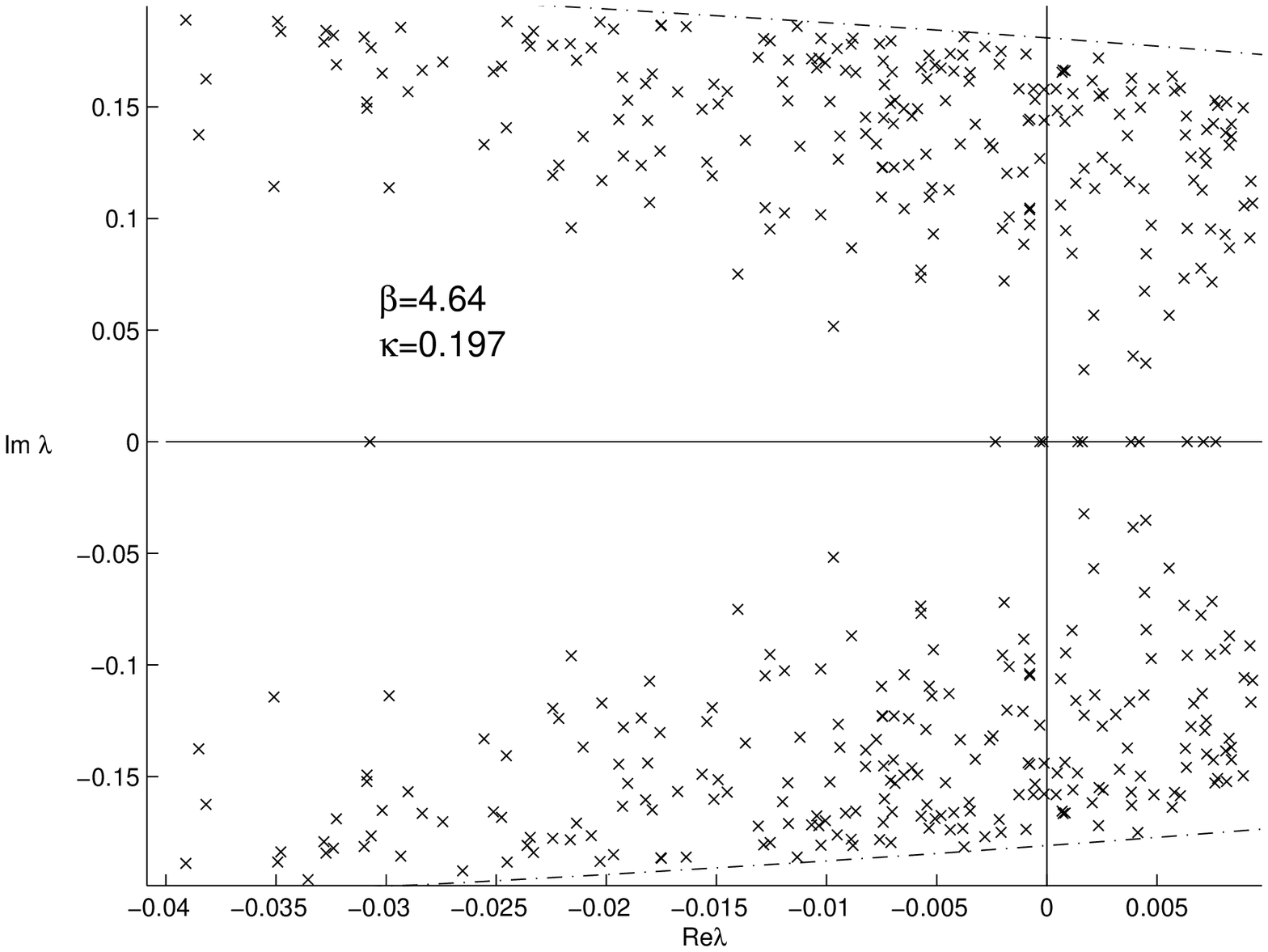}
\end{center}
\caption{
Low-lying eigenvalues for a set of 10 configurations with exceptionally
small eigenvalues, at $\beta=4.68$ and $\kappa=0.195$ (left panel), 
 $\beta=4.64$ and $\kappa=0.197$ (middle panel), detail  (right panel).
}
\label{sec4:fig:exc}
\end{figure*}

 In figure \ref{sec4:fig:exc} two significant examples
 are reported.
 The first set of configurations in the figure  corresponds to
 a moderately small quark mass (run $(h)$) and the second to a very small
 quark mass (run $(i)$).
 In both cases we selected the configurations with smallest eigenvalues
 of $\tilde{Q}^2$.
 Even in this way we could not find a single real negative eigenvalue
 for the first run $(h)$.
 In the second case we found three configurations with negative eigenvalues.
 The (in total) four negative eigenvalues are visible in the detail in the
 right panel of figure \ref{sec4:fig:exc}.
 As stated before these configurations are statistically insignificant.

 Some comments are in order.
 We have collected strong evidence that the presence of configurations
 with negative determinant is irrelevant at this stage. 
 Of course it is not yet possible to tell how this picture will evolve
 on larger volumes and closer to the continuum limit.
 It will be necessary to keep monitoring the low part of the
 spectrum as we did here.
 Since, to that purpose, we only need to know a very small part of the
 spectrum, there is no reason to think that this task should become too
 difficult on large volumes. 

 As a last remark we should stress that we performed this analysis of
 the sign for very small quark masses.
 Even if (partially quenched) Chiral Perturbation Theory is valid for
 any combination of the quark masses, it is probably not worth having
 an unpaired sea quark with a mass much smaller than the strange quark.
 Therefore, provided that the picture will not dramatically change on
 larger lattices, for all {\em physical} circumstances it seems very
 unlikely that the determinant sign could become a problem.

\subsection{Flow of eigenvalues}\label{sec4.3}

\begin{figure*}
\begin{center}
\includegraphics[height=0.4\textwidth,width=0.4\textwidth]{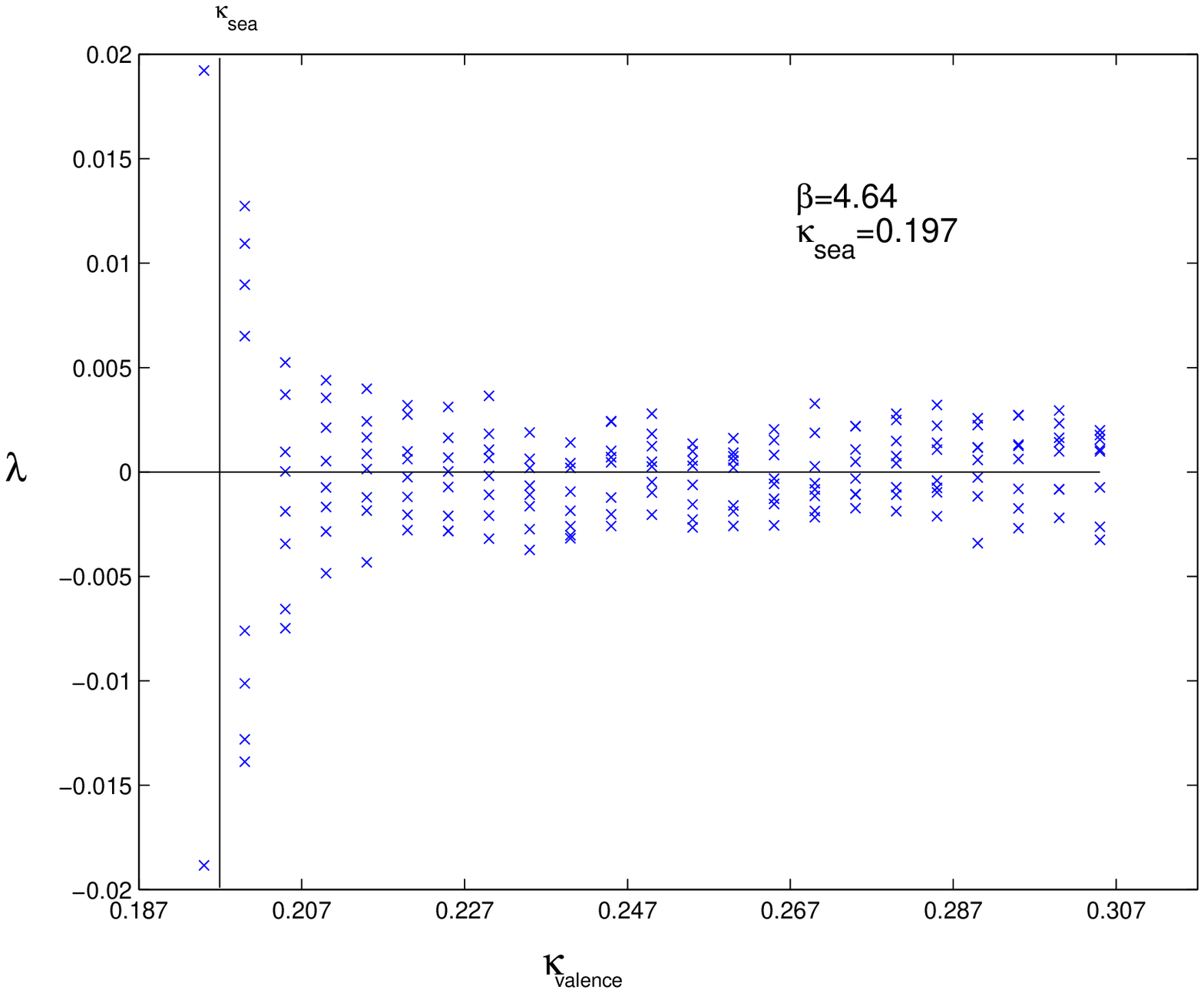}
\hspace{1cm}
\includegraphics[height=0.4\textwidth,width=0.4\textwidth]{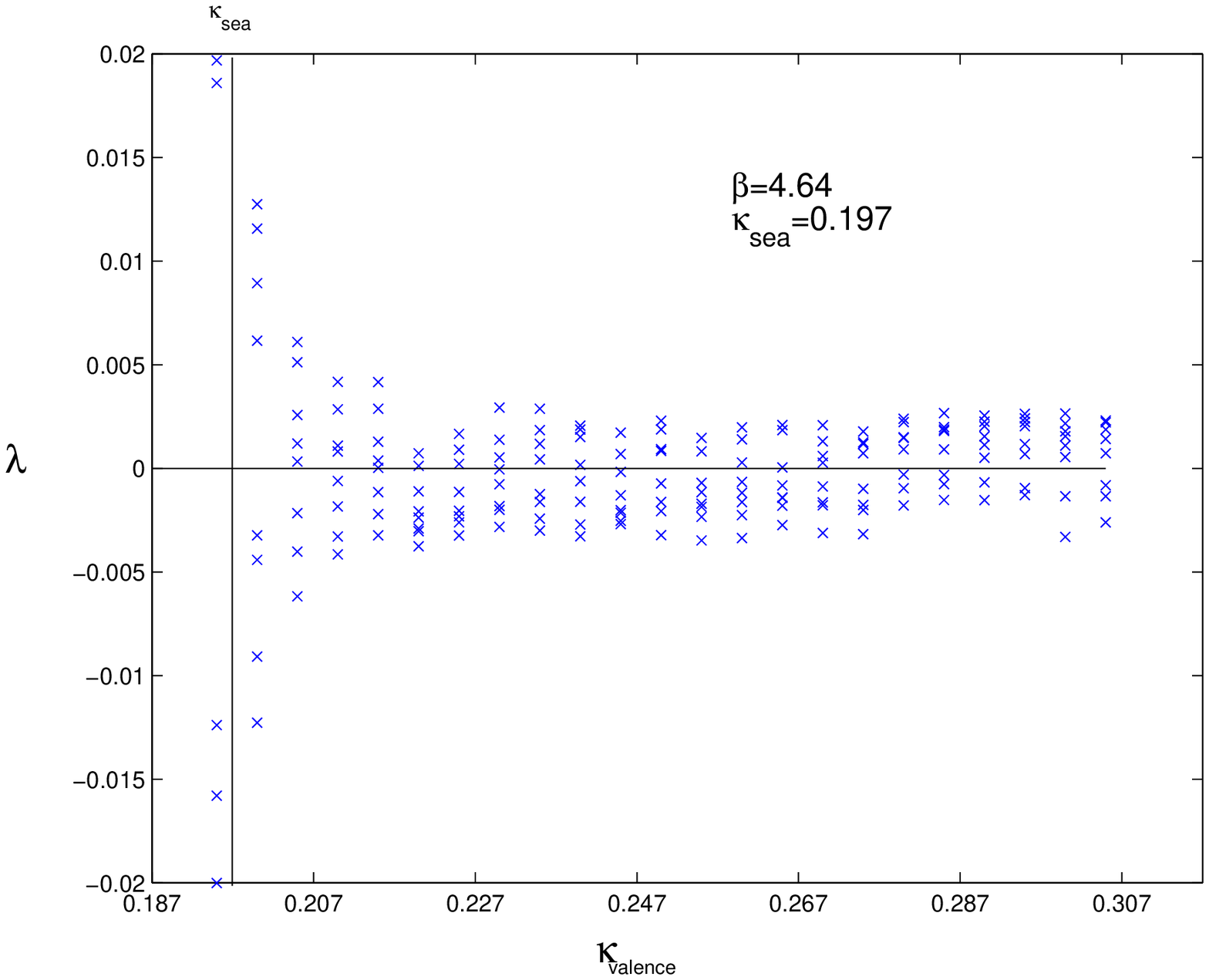}
\end{center}
\caption{
 Computation of 8 eigenvalues closest to zero of the hermitean
 Wilson-Dirac matrix for two configurations from the run at $\beta=4.64$ and
 $\kappa=0.197$.
}
\label{sec4:fig:NM}
\end{figure*}

 By using the algorithm of Kalkreuther and Simma \cite{Kalkreuter:SIMMA}
 we also explored the flow of the spectrum $\{\tilde{\lambda}\}$
 of the hermitean matrix $\tilde{Q}$ for a wide range of valence 
 $\kappa$ values, going from
 zero bare quark mass to a large negative one.
 This is interesting in view of simulations of dynamical fermions
 with Neuberger's operator \cite{Neuberger:OPE}, where the inverse square root
 of $\tilde{Q}^2$ with negative valence mass has to be taken.
 The optimal valence mass should be chosen in a region where
 $\tilde{Q}$ has no eigenvalues extremely close to zero,
 namely where a ``gap'' is opening up in the spectrum near
 $\tilde{\lambda}=0$.
 The results for two typical configurations are plotted in figure
 \ref{sec4:fig:NM}.
 For large negative masses we observed many sign changes and the
 eigenvalue with smallest absolute value is always close to zero.
 It seems that for dynamical Wilson fermions on our coarse lattice there
 is no gap-opening near $\tilde{\lambda}=0$.

 A possible application of the eigenvalue flow is to monitor the
 number of negative eigenvalues at $\kappa=\kappa_{sea}$
 \cite{Edwards:FLOW,Campos:GLUINO}.
 This is substantially cheaper than the analysis of the spectrum of
 non-hermitean matrix $Q$ by the Arnoldi method.
 For instance, observing the eigenvalue flow one can easily exclude
 the absence of negative eigenvalues if there is no crossing of zero
 in the flow below $\kappa_{sea}$ -- which is the typical case.
 A more detailed (and more expensive) analysis can be restricted to the
 seldom case when a crossing occurs.

\section{Discussion}\label{sec5}

 Our runs on $8^3 \cdot 16$ lattices with lattice spacing of about
 $a \simeq 0.27\,{\rm fm}$ for $N_f=2$ degenerate quarks display the
 dependence of simulation costs on the quark mass.
 Assuming the parametrization in (\ref{eq1.01}) with $z_L=5$ and
 $z_a=2$, from the integrated autocorrelation of the average plaquette
 we obtain
\begin{equation}\label{eq5.01}
z_\pi \simeq 4 \ , \hspace{3em}
F \simeq 0.8 \cdot 10^9\, {\rm flop} \ .
\end{equation}
 The power for the quark mass dependence $z_\pi$ comes out smaller than
 $z_{\pi\rho}=6$ in the form (\ref{eq1.02}) quoted by the CP-PACS, JLQCD
 Collaboration \cite{Ukawa:BERLIN} but if we omit the point with largest
 quark mass and perform a fit with the para\-metrization (\ref{eq1.02}) we
 also obtain $z_{\pi\rho} \simeq 6$ (see figure \ref{fig_tau}).

 As shown by figure \ref{fig_tau} (left panel), our data on the integrated
 autocorrelation of the average plaquette are well fitted by $z_\pi=4$
 in the whole range $0.6 \leq M_r \leq 6$ which approximately
 corresponds to $\frac{1}{5}m_s \leq m_{ud} \leq 2m_s$.
 The data on the integrated autocorrelation of the smallest eigenvalue
 of the squared hermitean fermion matrix show an even weaker power
 $z_\pi \simeq 3$ but there the errors are larger and the fit is less
 convincing (see right panel in figure \ref{fig_tau}).
 The pion mass has the shortest autocorrelation; this also shows a power
 $z_\pi\simeq 3$.
 $z_\pi=4$ corresponds to a behaviour proportional to the inverse square
 of the quark mass.
 Qualitatively speaking, according to table \ref{tab_auto}, one inverse
 quark mass power is due to the increase of the condition number of the
 fermion matrix and another inverse power comes from the increase of the
 autocorrelation in numbers of update cycles. Note that because of
 $(r_0 m_\pi)^{2} \propto (r_0 m_q)$ the case of
 $z_\pi = 4,\; z_a = 2$ corresponds to a situation when the scale
 parameter $r_0$ cancels in the cost formula (\ref{eq1.01}).

 The overall factor $F$ given in (\ref{eq5.01}) is such that for our
 second smallest quark mass $M_r \simeq 0.6$ (run $(i)$) 
 the cost in floating point
 operations comes out to be $C \simeq 2.3 \cdot 10^{14}$.
 As table \ref{tab_auto} shows, considering instead of the integrated
 autocorrelation of the average plaquette the one of the pion mass,
 the result is $C \simeq 0.4 \cdot 10^{14}$.
 The parameters in (\ref{eq1.02}) give the same number.
 The other estimates for Wilson-type quarks in \cite{Lippert:BERLIN} and
 \cite{Wittig:BERLIN} in this point are:
 $C_L \simeq 0.2 \cdot 10^{14}$ and $C_W \simeq 1.1 \cdot 10^{14}$,
 respectively.
 Taking into account that the numbers $C_{U,L,W}$ have been obtained
 under rather different circumstances concerning simulation algorithm,
 autocorrelations considered, quark mass range, lattice size and even
 lattice action there is a surprisingly good order of magnitude
 agreement.

\begin{figure}
\begin{center}
\includegraphics[angle=-90,width=\colw]{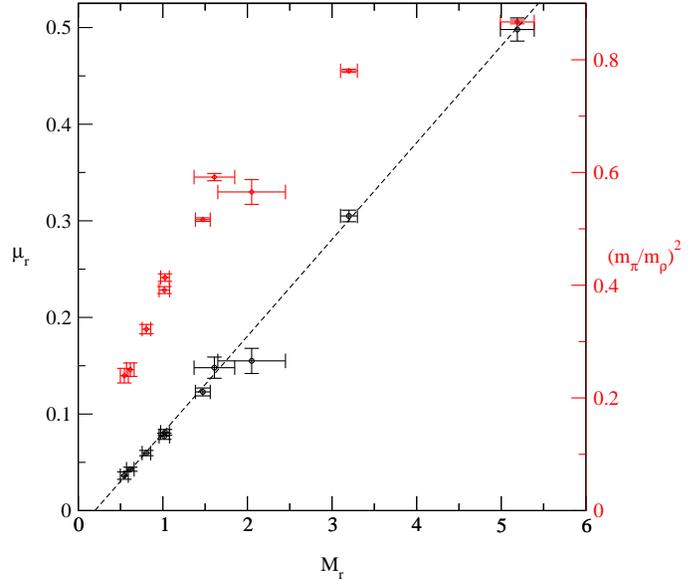}
\end{center}
\caption{ 
The dependence of $(m_\pi/m_\rho)^2$ on $M_r$ according to table
\protect{\ref{tab_resu}} (right values).
The values of $\mu_r$ are also shown (left) together with a linear
fit (dashed line).}
\label{sec5:fig:MFit}
\end{figure}

 It is remarkable that in a rather broad range of quark masses
 $\frac{1}{5}m_s \leq m_{ud} \leq m_s$ (leaving out the point at
 $m_{ud} \simeq 2m_s$) two fits with $z_\pi=4$ and $z_{\pi\rho}=6$
 work equally well (figure \ref{fig_tau}).
 This implies in this range of quark masses a
 peculiar dependence of $m_\pi/m_\rho$ on $M_r$ (see figure 
 \ref{sec5:fig:MFit},
 where the relation between the two different quark mass
 parameters $\mu_r$ and $M_r$ is also shown). However, the two
 parametrizations in (\ref{eq1.01}) and (\ref{eq1.02}) cannot
 be both correct in the vicinity of zero quark mass because 
 there the two powers have to be equal: $z_{\pi\rho}=z_\pi$.
 Saying it differently, the extrapolations of the two fits below
 $m_{ud}=\frac{1}{5}m_s$ are different.
 The fit with $z_{\pi\rho}=6$ gives a more dramatic slowing down near
 zero quark mass than the one with $z_\pi=4$.
 The real asymptotics near $m_{ud}=0$ could be disentangled by going to
 still smaller quark masses.
 With TSMB there is no serious obstacle for doing this -- except for the
 increase in necessary computer time.

 The value of the lattice spacing in this paper is chosen rather large
 in order to limit the computational costs for these tests.
 Our aim was to concentrate on the quark mass dependence in the range
 of light quarks.
 Further studies will be needed for investigating the cost as a
 function of the lattice spacing (in particular, the value of the
 exponent $z_a$) for smaller values of $a$.
 In this respect the experience of the DESY-M\"unster-Roma
 Collaboration in the supersymmetric Yang-Mills theory at much smaller
 lattice spacings ($a \simeq 0.06 {\rm fm}$)
 \cite{Campos:GLUINO,Farchioni:SUSYWTI} shows already that TSMB has a
 decent behaviour also closer to the continuum limit.

 Besides the quark mass dependence of simulation costs, the other
 interesting question we investigated in this paper is the
 distribution of the small eigenvalues of the fermion matrix and, in
 particular, the existence of negative fermion determinants of a single
 quark flavour.
 Our data show that the effect of the fermion determinant is rather
 explicit because of the strong suppression of the eigenvalue density
 near zero (see figures \ref{sec4:fig:sys:atoj}-\ref{sec4:fig:exc}).
 The statistical weight of configurations with negative determinant
 is negligible even at our smallest quark masses.
 In fact after an extensive analysis we only found three configurations 
 with negative determinant at our second smallest quark mass $M_r \simeq 0.6$ 
 (run $(i)$) and none of them at other quark masses.
 Taking into account the small reweighting factors of the configurations
 with negative determinant, their relative statistical weight is 
 ${\cal O}(10^{-5})$.

 It is clear that it would be important to check the volume dependence
 of our results, both for simulation costs and small eigenvalues,
 on larger lattices and closer to the continuum limit.
 We plan to do this in the future.

$\phantom{x}$

{\large\bf Acknowledgement}

\noindent
 The computations were performed on the APEmille systems installed 
 at NIC Zeuthen, the Cray T3E systems at NIC J\"ulich and the PC
 clusters at DESY Hamburg.
 We thank H.~Simma for help with APEmille programming and optimization,
 M.~L\"uscher for providing his SSE code, H.~Wittig for useful
 discussions on various topics, T.~Kovacs and W.~Schroers for sharing with
 us their experience about the computation of eigenvalues, and R. Peetz for
 careful reading of the manuscript.

\bibliography{}

\end{document}